\documentclass[twocolumn,twocolappendix]{aastex63}

\graphicspath{{./}{figures/}}
\usepackage[nolist]{acronym}
\newacro{adi}[ADI]{angular differential imaging}
\newacro{ao}[AO]{adaptive optics}
\newacro{aolp}[AOLP]{angle of linear polarization}
\newacro{bff}[BFF]{Best Factor Finding}
\newacro{charis}[CHARIS]{Coronagraphic High Angular Resolution Imaging Spectrograph}
\newacro{css}[CSS]{circumstellar signal}
\newacro{de}[DE]{differential evolution}
\newacro{disnmf}[DI-sNMF]{data imputation using sequential nonnegative matrix factorization}
\newacro{dpp}[DPP]{Data Processing Pipeline}
\newacro{drp}[DRP]{Data Reduction Pipeline}
\newacro{fov}[FOV]{field of view}
\newacro{fwhm}[FWHM]{full width at half maximum}
\newacro{gpi}[GPI]{Gemini Planet Imager}
\newacro{gto}[GTO]{Guaranteed Time Observations}
\newacro{hiciao}[HiCIAO]{High-Contrast Coronographic Imager for Adaptive Optics}
\newacro{hst}[HST]{Hubble Space Telescope}
\newacro{hwp}[HWP]{half-wave plate}
\newacro{i}[I]{total intensity}
\newacro{ifs}[IFS]{integral field spectrograph}
\newacro{irdis}[IRDIS]{InfraRed Dual-band Imager and Spectrograph}
\newacro{jwst}[JWST]{the James Webb Space Telescope}
\newacro{klip}[KLIP]{Karhunen-Lo\`{e}ve Image Projection}
\newacro{loci}[LOCI]{Locally Optimized Combination of Images}
\newacro{magaox}[MagAO-X]{MagAO-X}
\newacro{mloci}[MLOCI]{Matched LOCI}
\newacro{nir}[NIR]{near-infrared}
\newacro{nmf}[NMF]{Non-negative Matrix Factorization}
\newacro{pca}[PCA]{principal component analysis}
\newacro{pdi}[PDI]{polarimetric differential imaging}
\newacro{pi}[PI]{polarized intensity}
\newacro{piaacmc}[PIAACMC]{phase-induced amplitude apodization complex mask coronagraph}
\newacro{ppd}[PPD]{protoplanetary disk}
\newacro{psf}[PSF]{point spread function}
\newacro{rdi}[RDI]{reference star differential imaging}
\newacro{sb}[SB]{surface brightness}
\newacro{scexao}[SCExAO]{Subaru  Coronagraphic  Extreme  Adaptive  Optics}
\newacro{sdi}[SDI]{spectral differential imaging}
\newacro{snr}[SNR]{signal-to-noise ratio}
\newacro{snre}[SNRE]{signal-to-noise per resolution element}
\newacro{spf}[SPF]{scattering phase function}
\newacro{sphere}[SPHERE]{Spectro-Polarimetric High-contrast Exoplanet REsearch instrument}
\newacro{vampires}[VAMPIRES]{Visible Aperture Masking Polarimetric Imager for Resolved Exoplanetary Structures}
\newacro{vapp}[vAPP]{vector Apodizing Phase Plate}
\newacro{vlt}[VLT]{Very Large Telescope}
\usepackage{breqn}

\newcommand{\citeac}[2]{\aclu{#1} \citep[\ac{#1};][]{#2}}

\begin{document}
\title{Constrained Reference Star Differential Imaging: Enabling High-Fidelity Imagery of Highly Structured Circumstellar Disks\footnote{Based on data collected at Subaru Telescope, which is operated by the National Astronomical Observatory of Japan.}}

\correspondingauthor{Kellen Lawson}
\email{kellenlawson@gmail.com}
\author[0000-0002-6964-8732]{Kellen Lawson}
\affiliation{Department of Physics and Astronomy, University of Oklahoma, Norman, OK}

\author[0000-0002-7405-3119]{Thayne Currie}
\affiliation{Subaru Telescope, National Astronomical Observatory of Japan, 
650 North A`oh$\bar{o}$k$\bar{u}$ Place, Hilo, HI  96720, USA}
\affiliation{Department of Physics and Astronomy, University of Texas-San Antonio, 1 UTSA Circle, San Antonio, TX, USA}
\affiliation{Eureka Scientific, 2452 Delmer Street Suite 100, Oakland, CA, USA}

\author[0000-0001-9209-1808]{John P. Wisniewski}
\affiliation{Department of Physics and Astronomy, University of Oklahoma, Norman, OK}

\author[0000-0001-5978-3247]{Tyler D. Groff}
\affiliation{NASA-Goddard Space Flight Center, Greenbelt, MD, USA}

\author[0000-0003-0241-8956]{Michael W. McElwain}
\affiliation{NASA-Goddard Space Flight Center, Greenbelt, MD, USA}

\author[0000-0001-5347-7062]{Joshua E. Schlieder}
\affiliation{NASA-Goddard Space Flight Center, Greenbelt, MD, USA}

\submitjournal{ApJL}
\accepted{2022 July 29}

\begin{abstract}
High-contrast imaging 
presents us with the opportunity to study circumstellar disks and the planets still embedded within them --- providing key insights into the formation and evolution of planetary systems. However, the post-processing techniques that are often needed to suppress stellar halo light typically result in significant and variable loss of circumstellar light, even when using relatively conservative approaches like \ac{rdi}.
We introduce ``constrained reference star differential imaging" (constrained RDI), a new class of \ac{rdi} \ac{psf}-subtraction techniques for systems with circumstellar disks.  Constrained RDI utilizes either high-resolution \ac{pi} images or disk models to severely limit or even eliminate the signal loss due to oversubtraction that is common to \ac{rdi}. 

We demonstrate the ability of constrained \ac{rdi} utilizing polarimetric data to yield an oversubtraction-free detection of the AB Aurigae protoplanetary disk in total intensity. \Ac{pi}-constrained \ac{rdi} allows us to decisively recover the spectral signature of the confirmed, recently-discovered protoplanet, AB Aurigae b (Currie et al. 2022).   We further demonstrate that constrained \ac{rdi} can be a powerful analysis tool for soon-to-be-acquired James Webb Space Telescope coronagraphic imaging of disks.   In both cases, constrained \ac{rdi} provides analysis-ready products that enable more detailed studies of disks and more robust verification of embedded exoplanets.
\end{abstract}

\section{Introduction}\acresetall
Circumstellar disk systems serve as benchmarks for the study of how and where exoplanets form. With the advent of ground-based extreme \ac{ao} facilities, such as Subaru's \acsu{scexao} \citep{Jovanovic2015}, VLT's \acsu{sphere} \citep{Beuzit2019}, Gemini's \acsu{gpi} \citep{Macintosh2015}, and Magellan's \acsu{magaox} \citep{Males2018}, and with the recent launch of \ac{jwst}, high-contrast imaging studies of disks have reached an exciting new era. We can now spatially resolve the morphological signatures within disks (e.g., gaps or spirals) thought to be caused by newly formed/forming companions while also directly identifying \& characterizing the young planets and sub-stellar objects that may cause them \citep[e.g.,][]{Keppler2018}. Combined with multi-wavelength or \ac{ifs} data, we have the technology to conduct incredibly detailed spatial and spectral analysis of these systems.

\begin{figure}
\centering
\includegraphics[width=0.475\textwidth]{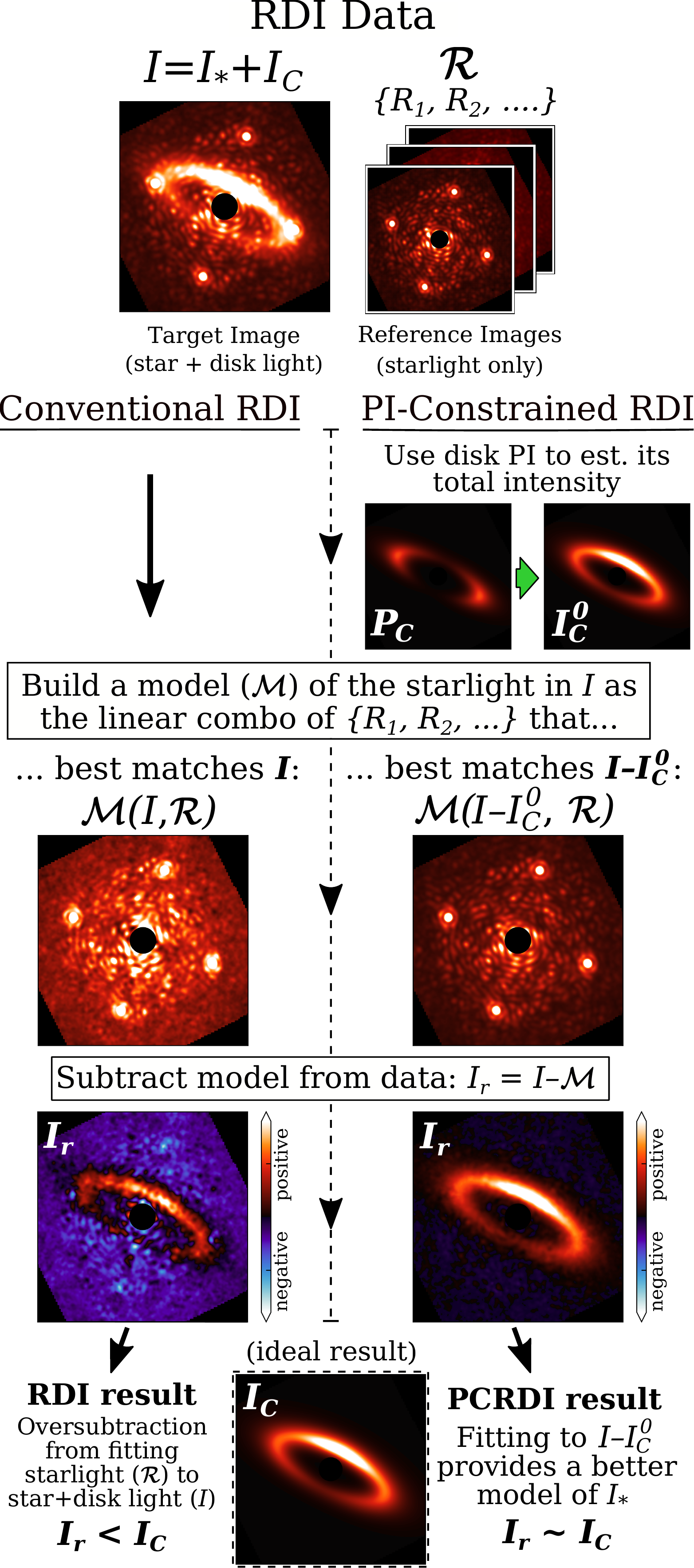}
\caption{\textbf{Top center}: \Ac{rdi} data for a (synthetic) disk system. \textbf{Left column}: Conventional \ac{rdi} \ac{psf} subtraction, yielding significantly attenuated results due to oversubtraction (``RDI result", lower left). \textbf{Right column}: PI-Constrained RDI (PCRDI; Section \ref{sec:pcrdi}). The PCRDI result (lower right) avoids attenuation by using \ac{pi} disk imagery to estimate and suppress disk flux while modeling the diffraction pattern of the starlight in total intensity. \label{fig:rdi_pcrdi_explanation}}
\end{figure}

A key challenge for these studies is the isolation of \ac{css} from the bright pattern of diffracted starlight from the host star (the stellar \aclu{psf} or \acsu{psf}). For disk-focused studies, this is commonly achieved using \ac{rdi}. ``Classical \ac{rdi}", in which a reference star image is directly subtracted from the target image, can leave significant residual starlight where the \ac{psf} changes significantly between exposures (e.g., at narrow separations). More advanced techniques --- 
e.g., \citeac{loci}{Lafreniere2007} or \citeac{klip}{Soummer2012} 
--- better model the starlight but also cause some \ac{css} to be lost (or ``attenuated''; see Fig. \ref{fig:rdi_pcrdi_explanation}, top row). Since this \ac{css} loss is neither spatially nor spectrally (for multi-wavelength data) uniform \citep[e.g., Figure 2 in ][]{Betti2022}, features identified within disks  -- including planets -- can be challenging to validate. \Ac{pdi} complements conventional \ac{i} disk studies by producing unattenuated \ac{pi} imagery of disks. However, since emission from young planets is not expected to be significantly polarized, planets cannot generally be detected in \ac{pi} alone. Though the comparison of \ac{i} and \ac{pi} should highlight planets due to their diminished polarization relative to highly polarized disk material, the non-uniform loss of signal in total intensity precludes this measurement. 

In some scenarios, the challenges presented by disk signal loss and the confusion of disk and planet signal can be circumvented using forward modeling techniques \citep[e.g.,][]{Currie2015,Pueyo2016,Mazoyer2020, Lawson2020}. In disk forward modeling, signal loss from \ac{psf}-subtraction is induced on synthetic disk images of varying parameters until the observed (attenuated) \ac{css} is reasonably reproduced. This enables robust assessment of disk geometry in the presence of signal loss, while also allowing approximate correction of disk surface brightness measurements for some parts of the disk \citep[e.g.,][]{Bhowmik2019, Betti2022}. However, as this requires generating and processing models spanning the breadth of plausible disk parameters \citep{Pueyo2016}, it is generally infeasible for structurally complicated disks (e.g., having spiral arms) for which more model parameters must be explored and for which individual models can be more expensive to generate. In the case of simpler disks where exhaustive modeling is feasible, it nevertheless introduces intractable uncertainties for analysis and is often a significant computational bottleneck.

In recent years, an array of advancements for \ac{psf}-subtraction --- leveraging \ac{rdi} as well as \ac{adi} and \ac{sdi} --- have been developed for the purpose of exoplanet and disk imaging. These include techniques which focus on more thoroughly removing starlight and/or better recovering point sources \citep[e.g., \textit{TLOCI}, \textit{A-LOCI}, \textit{PACO};][]{Marois2014, Currie2012, Flasseur2018,Currie2022b}, but which still face steep challenges in addressing the attenuation of extended \ac{css} through forward-modeling. More relevant for disk studies, \ac{nmf} with the \ac{bff} procedure \citep{Ren2018} and \citeac{disnmf}{Ren2020} are intended to mitigate \ac{css} loss for disk systems. However, they depend on the availability of regions free of \ac{css} to completely eliminate signal loss and are thus less effective for {more extended disks or disks which dominate small separations (where temporal \ac{psf} changes are most significant)}. More recently, {``source-separation" algorithms, such as \emph{MAYONNAISE} \citep{Pairet2021} and \emph{REXPACO} \citep{Flasseur2021}, have provided sophisticated tools for isolating \ac{css} and limiting the negative effects of \ac{psf}-subtraction. Source-separation techniques attempt to simultaneously model each significant source of signal throughout a data sequence --- e.g., separately considering the starlight and any rotating (in an ADI sequence) \ac{css}, as well as the time-variable \ac{psf}, the coronagraph, and noise. As both tools significantly leverage the continuous spatial+temporal variations throughout an \ac{adi} (or ADI+SDI) sequence to disentangle sources of light, it is unclear if they will be effective for space-based observations, such as those from \ac{jwst}, for which disk targets are typically observed at only two distinct roll-angles. 
Further, while they are intended to precisely isolate \ac{css}, they do not provide any clear methods for assessing, quantifying, or correcting any inaccuracy that might remain. As it stands, d}espite the exceptional capabilities of current and upcoming observatories, our ability to study circumstellar disks (and the young planets within them) is inhibited by the loss of \ac{css} signal during post-processing and the difficulty of quantifying this loss.

In this work, we describe a new class of \ac{rdi} \ac{psf}-subtraction techniques for circumstellar disk systems: constrained \ac{rdi}. In constrained \ac{rdi}, available information regarding a disk (e.g., through polarimetry) is used to prevent \ac{css} signal loss during \ac{psf}-subtraction. By tuning constraints to best explain a target's observations, constrained \ac{rdi} can be optimized to effectively eliminate \ac{css} loss during \ac{psf}-subtraction.  We generalize these techniques for use with any \ac{rdi}-based \ac{psf}-subtraction technique in which reference images are combined in some manner to minimize residuals with the target data (e.g., LOCI, KLIP, NMF, etc.; Sections \ref{sec:rdi_attenuation}, \ref{sec:constrained_rdi}). 

Using \ac{pi}-based constraints, we demonstrate constrained \ac{rdi} in application to simulated \ac{charis} \ac{ifs} observations of a spiral armed disk system (Section \ref{sec:sims}) and to on-sky \ac{charis} observations of the AB Aurigae protoplanetary disk system --- whose embedded protoplanet is verified by this approach \citep[Section \ref{sec:abaur};][]{Currie2022}. Additionally, using synthetic disk models as constraints, we apply constrained \ac{rdi} to simulated \ac{jwst} NIRCam observations (Section \ref{sec:mcrdi}) to demonstrate the approach's efficacy for upcoming JWST observations. These techniques are broadly applicable for nearly any disk imaging study, including those using data from: ground-based observatories, \ac{hst}, \ac{jwst}, and future observatories (e.g., Roman Space Telescope, 30-meter class telescopes, and others; Section \ref{sec:broader_apps}). Moreover, constrained \ac{rdi} can be implemented in existing pipelines with only minor changes ({Section \ref{sec:broader_apps}}), is extremely unlikely to induce spurious (false-positive) circumstellar features ({Section \ref{sec:limitations}}){, and its results can be assessed using standard methods familiar to the disk imaging community (e.g., forward modeling; Appendix \ref{app:pcrdi_fwdmod}).}

\section{Disk Signal Attenuation in Reference Star Differential Imaging}\label{sec:rdi_attenuation}

Consider a (total intensity) target image, $I$, containing both stellar signal, $I_*$, and circumstellar signal, $I_{C}$, such that $I=I_* + I_{C}$. The objective of \ac{psf}-subtraction is to determine the stellar signal in the image so that it can be subtracted from the target frame to isolate the \acf{css}. In \acf{rdi}, this problem is addressed by using observations of an additional star for which no \ac{css} is present. To accommodate temporal changes to the \ac{psf} between the target and reference frames, it is common to utilize a set of reference images from which an optimal match to each target frame can be determined \citep[e.g.,][]{Lafreniere2007, Soummer2012, Choquet2014, Hagan2018}.

Let $\mathcal{R}$ be a sequence of $N$ reference star images containing only stellar signal: $\mathcal{R} = \{ R_1, R_2, \ldots, R_N \}$. Let $\mathcal{M}(I, \mathcal{R})$ be the \ac{psf} model constructed from the reference data $\mathcal{R}$ to minimize residuals with a target image $I$.  $\mathcal{M}(I, \mathcal{R})$ could be a \ac{psf} model constructed from a linear combination of reference frames or with a \ac{pca} based technique such as KLIP.

For conventional RDI \ac{psf}-subtraction techniques (hereafter, simply referred to as ``RDI"), the residuals, $I_{r}$, are determined as\footnote{{For simplicity, notation herein refers to a single target image. In application, there are generally multiple such images which would each be PSF-subtracted in the same manner before being derotated to a north-up orientation and averaged together to form the final result.}}:

\begin{equation}\label{eq:rdi_res}
I_{r} = I - \mathcal{M}(I, \mathcal{R})
\end{equation}

In an ideal scenario, the \ac{psf} model would perfectly reproduce the stellar signal in the target image, $\mathcal{M}(I, \mathcal{R}) = I_*$, and so the residuals would simply be the \ac{css} in the target image: $I_{r} = I - I_* = I_C$. In practice, the presence of \ac{css} in $I$ directly influences the \ac{psf} model that is constructed; rather than constructing the stellar \ac{psf} model that minimizes the residuals with the target stellar \ac{psf}, we actually construct the model that minimizes the residuals with the stellar \emph{and} circumstellar signal. This results in a \ac{psf} model that is brighter than the starlight in the target image, and thus the circumstellar signal in the residuals is artificially reduced in brightness: $I_{r} < I_{C}$. This is referred to as ``oversubtraction" and is the sole source of systematic attenuation for \ac{rdi} \citep{Pueyo2016}. In this notation, the attenuated \ac{css} result, $I_C^\prime$, for a given $I_C$ is found simply by replacing $I$ with $I_C$ in Eq. \ref{eq:rdi_res}:

\begin{equation}\label{eq:rdi_atten}
I_C^\prime = I_C - \mathcal{M}(I_C, \mathcal{R})
\end{equation}
For point-like companions, this effect is generally minor. For extended \ac{css}, such as from a circumstellar disk, the effect can be severe \citep[e.g.,][]{Lawson2021b, Betti2022}.

\section{Mitigating RDI Oversubtraction: Constrained RDI}\label{sec:constrained_rdi}
If the underlying circumstellar signal in $I$, $I_{C}$, was known \textit{a priori}, oversubtraction could be eliminated by computing the residuals as: $I_{r} = I - \mathcal{M}(I-I_{C}, \mathcal{R})$, or in other words, $I_{r} = I - \mathcal{M}(I_*, \mathcal{R})$. Of course, since $I_C$ is the desired product of \ac{rdi}, this provides no immediate utility. 

However, we can approximate $I_C$ to mitigate oversubtraction in \ac{rdi} products. Denoting an estimate of $I_C$ as $I_{C}^{0}$, the residuals are then:
\begin{equation}\label{eq:pcrdi_res}
I_{r} = I - \mathcal{M}(I-I_{C}^{0}, \mathcal{R})
\end{equation}

{Notably, as in standard \ac{rdi} reductions, the \ac{psf} model itself is still constructed entirely from reference images (or the eigenvectors of the reference image covariance matrix in the case of KLIP). The estimate of $I_C$, $I_{C}^{0}$, is used exclusively in determining the optimal combination of reference images to use in the \ac{psf} model and is not directly used in the final product; the effect of the constraint must always be some combination of reference images.}
{This provides substantial insulation against the introduction of false-positive circumstellar features. A particularly poor estimate of the circumstellar signal contained in the data could result in worse \ac{psf}-subtraction (e.g., by causing under-subtraction). However, to induce spurious features from the estimate in the final result, such features would need to be reproduced by some combination of the reference images and at a sufficient number of parallactic angles to remain after the sequence is derotated and averaged. For data with non-negligible field rotation, using small/truncated reference sequences (e.g, retaining the most correlated images from a larger reference library), and using large/full-frame \ac{psf} subtraction regions, this is very unlikely (see Section \ref{sec:limitations} for further discussion). Moreover, as with conventional RDI reductions, any features that do manifest can be verified using forward modeling techniques (Appendix \ref{app:pcrdi_fwdmod}). As such, this approach provides an extremely safe means by which to improve \ac{rdi} reductions of circumstellar disk targets.}

\subsection{Polarized Intensity Constrained RDI}\label{sec:pcrdi}
For targets also observed with \acf{pdi}, the unattenuated \acf{pi} of circumstellar signal, $P_C$, can be attained using standard techniques \citep[e.g., double-differencing;][]{Kuhn2001}. Given that, by definition, $I_C \geq P_C$, \ac{rdi} results can be improved in such a case by adopting $I_{C}^{0} = P_{C}$ and proceeding as in Eq. \ref{eq:pcrdi_res}. Generally, using $P_C$ as a conservative estimate of $I_{C}$ will offer an unambiguous improvement for \ac{rdi} products -- though the improvement may be small for particularly compact disks (where oversubtraction was already minor) or disks with low fractional polarization (where $I_C \gg P_C$). 

However, we can do much better by using reasonable assumptions regarding the scattering properties of circumstellar material. As a function of scattering angle, $\phi$, the ratio of polarized to total intensity, or fractional polarization ($F_{pol}$), for scattered light is often described by Rayleigh polarization \citep[e.g.,][]{Whitney2013,  Stolker2016, Gonzalez2017}:
\begin{equation}\label{eq:rayleigh_pol}
\frac{\rm PI}{\rm I} = F_{pol} = \frac{1-\cos^2\phi}{1+\cos^2\phi}
\end{equation}
By assuming a particular scattering surface for the disk \citep[e.g., as in \textit{diskmap};][]{Stolker2016}, a map of the corresponding scattering angles probed throughout the field, $\Phi$, and thus the fractional polarization, can be derived. Inverting Eq. \ref{eq:rayleigh_pol}, this can be used to transform $P_C$ to an estimate of $I_C$:
\begin{equation}\label{eq:pcrdi_est}
I_C^0 = P_C \cdot \left(\frac{1+\cos^2\Phi}{1-\cos^2\Phi}\right) = \frac{P_C}{F_{pol}^0},
\end{equation}
where $F_{pol}^0$ denotes the estimate of the fractional polarization for a particular assumed surface geometry. This total intensity \ac{css} estimate, $I_C^0$, can then be used with Eq. \ref{eq:pcrdi_res} to carry out constrained RDI \ac{psf} subtraction. This process is visualized in the bottom row of Figure \ref{fig:rdi_pcrdi_explanation}. Hereafter, we refer to this as \ac{pi}-constrained RDI or PCRDI.

Using \textit{diskmap}, $P_C$ is transformed to $I_{C}^{0}$ by assuming a smooth scattering surface with a radial profile defined by $a$, $b$, and $c$ as $h(r) = a+br^c$, with a particular peak fractional polarization ($s$), and which is viewed at a particular orientation (inclination, $i$, and position angle, PA). For well-studied disks, disk modeling results from literature can be used to estimate these parameters and thus enable PCRDI. Alternatively, their values can be directly optimized (see Section \ref{sec:optimization}). See Section \ref{sec:limitations} for discussion of additional considerations.

\subsection{Model-Constrained RDI}\label{sec:mcrdi}
For disks with simple geometries but which lack suitable \ac{pi} imagery to enable PCRDI, synthetic disk models can be adopted as $I_C^0$ in place of the PI-based estimates used in PCRDI. We refer to this approach as Model Constrained RDI or MCRDI. If a suitable literature-based model is not available, the model's parameters can be optimized directly, much as for PCRDI (Section \ref{sec:optimization}). We note that this may be significantly more time consuming than for PCRDI; not only does generating the \ac{css} estimate for a particular set of parameters take longer (i.e., constructing an entire disk model for MCRDI versus doing some geometry for PCRDI), but there are also more parameters to consider. Moreover, these parameters can have significant degeneracies and local minima that may motivate a more thorough exploration of the parameter space if decent estimates of the parameters for the disk are not known \textit{a priori}. 

In general, the timescales required to reach an optimal MCRDI solution will be comparable to those required to optimize a disk model using conventional forward-modeling techniques \citep[e.g.,][]{Lawson2021b}. Nevertheless, this technique provides significant utility over forward-modeling. While an attenuated model in good agreement with the data can provide {approximate} attenuation corrections \citep[e.g.,][]{Bhowmik2019}, such corrections {become extremely noisy in regions where the processed model approaches zero and also scale any noise or residual starlight when correcting \ac{css} (meaning signal-to-noise is not improved this way; see Section \ref{sec:pcrdi_sims}). Using the same model to carry out MCRDI instead provides the same benefits but avoids inflating stellar residuals and noise --- enabling accurate analysis over more of the \ac{fov} and potentially revealing fine morphological features and faint extended signal that would otherwise not be recovered.}

\subsection{Optimization of Constrained RDI}\label{sec:optimization}
For a constrained RDI technique, the values of the parameters governing the \ac{css} estimate (e.g., $s, a, b, c, i,$ and $\rm{PA}$ for PCRDI) can be directly optimized to identify the best match to the true underlying \ac{css}. For this purpose, we compute the following:
\begin{equation}\label{eq:opt_pcrdi_obj}
y = (I-I_{C}^{0}) - \mathcal{M}(I-I_{C}^{0}, \mathcal{R})
\end{equation}
This is simply the conventional \ac{rdi} residual calculation (Eq. \ref{eq:rdi_res}) where $I$ is replaced by $I-I_{C}^{0}$ throughout. As our models of the stellar and circumstellar signal approach the true stellar and circumstellar signal, the residual signal contained in the image $y$ will generally decrease:
\begin{equation}
\begin{split}
\text{As }\mathcal{M}(I-I_{C}^{0}, \mathcal{R}) \to I_* \text{ and } I_{C}^{0} \to I_C, \\
y \to (I-I_C) - I_* =  I_* - I_* = \hat{0}
\end{split}
\end{equation}

Since our stellar model will also generally improve as our estimate of the circumstellar signal improves, $y$ can be used to assess the quality of our \ac{css} estimate, $I_C^0$. A straight-forward objective function for optimization is then simply the sum of the squares of these residuals binned to the spatial resolution of the data\footnote{In some contexts, it may be preferable to compute a $\chi^2$ metric using pixel-wise uncertainties -- as is common for disk modeling purposes \citep[e.g.,][]{Thalmann2013, Currie2019}. However, since the most common \ac{psf}-subtraction algorithms are solving some version of an unweighted least-squares problem, the unweighted objective function tends to converge more smoothly.}. This objective function can then be combined with an optimization algorithm of choice to automatically determine the optimal estimate of \ac{css}.
{This procedure might be thought of as an extension of the negative PSF imputation technique for point-sources introduced by \citet{Marois2010a}, in that both use the \ac{css}- and PSF-subtracted residuals of a data sequence to gauge the quality of a \ac{css} model.}

\section{Throughput Assessment using Simulated Data}\label{sec:sims}
\subsection{PCRDI Throughput}\label{sec:pcrdi_sims}
To test the throughput of PCRDI in application to CHARIS IFS data \citep{Groff2016}, we generated multi-wavelength synthetic disk image cubes in Stokes Q, U, and I using HO-CHUNK 3-D \citep{Whitney2013} for a spiral-armed disk of similar morphology to the disk of AB Aurigae (see Section \ref{sec:abaur}). To simulate an RDI sequence for this model disk, we divided a large CHARIS reference star sequence into two sets -- one of which will have the \ac{css} added (the sequence of ``target" images: $\mathcal{I} = \{ I_1, I_2, \ldots\}$), and the other which will be left unchanged (the ``reference" sequence, $\mathcal{R}$)\footnote{For more information regarding the size of the target and reference sequences, see Table \ref{tab:dataruntimes}.}. We then scaled the wavelength slices of the target sequence to match the brightness of AB Aur. The disk model cubes --  Q, U, and I -- were convolved with the \ac{psf} for the target sequence, which was determined using the average shape of the calibration satellite spots over the full sequence\footnote{PSF-convolution of the model cubes before rotation to the parallactic angles needed for the synthetic data sequence assumes a predominantly azimuthally symmetric PSF. This is a reasonable assumption for the \ac{charis} \ac{psf}, but is less reasonable for systems such as JWST NIRCam.}. After convolution, Q and U were combined to form the disk \ac{pi}, and the slices of the PI and I cubes ($P_C$ and $I_C$, respectively) were scaled to approximately match the brightness of AB Aurigae’s disk. Finally, the $I_C$ cube was rotated to the parallactic angle of each exposure in the target sequence, $\mathcal{I}$, and added to the target cubes. We note here that the details of the disk-to-star contrast and how the initial reference star sequence is divided to form the simulated target and reference sequences will affect the relative residual noise level and thus the \ac{snre} in the products, but will not affect the typical percentage of \ac{css} attenuated by oversubtraction. Similarly, while the quality of the spectral match between the target and the reference star in this simulated sequence is unrealistic (given that they are the same star), a) this should again affect only the residual noise rather than the attenuation, and b) the effect would generally be minor in application to IFS data anyway (given that the \ac{psf} model is optimized separately for each narrow wavelength channel).

For \ac{psf}-subtraction, the \ac{psf} model for each target frame is constructed using a linear combination of the reference frames. To accelerate the \ac{psf} subtraction procedure, images are compared within a single annular optimization region spanning $r= 10-28$ pixels ($\sim 0\farcs16-0\farcs45$) for all 22 wavelength channels. This optimization region is also used for calculation of the goodness of fit for each trial result (Eq. \ref{eq:opt_pcrdi_obj}). Allowing all six scattering surface parameters to vary in wide ranges, we conducted optimization using the Levenberg–Marquardt (or ``damped least-squares") algorithm \citep[as implemented in the Python package \textit{LMFIT}; ][]{Newville2014}.

\begin{figure*}
    \centering
    \includegraphics[width=0.9\textwidth]{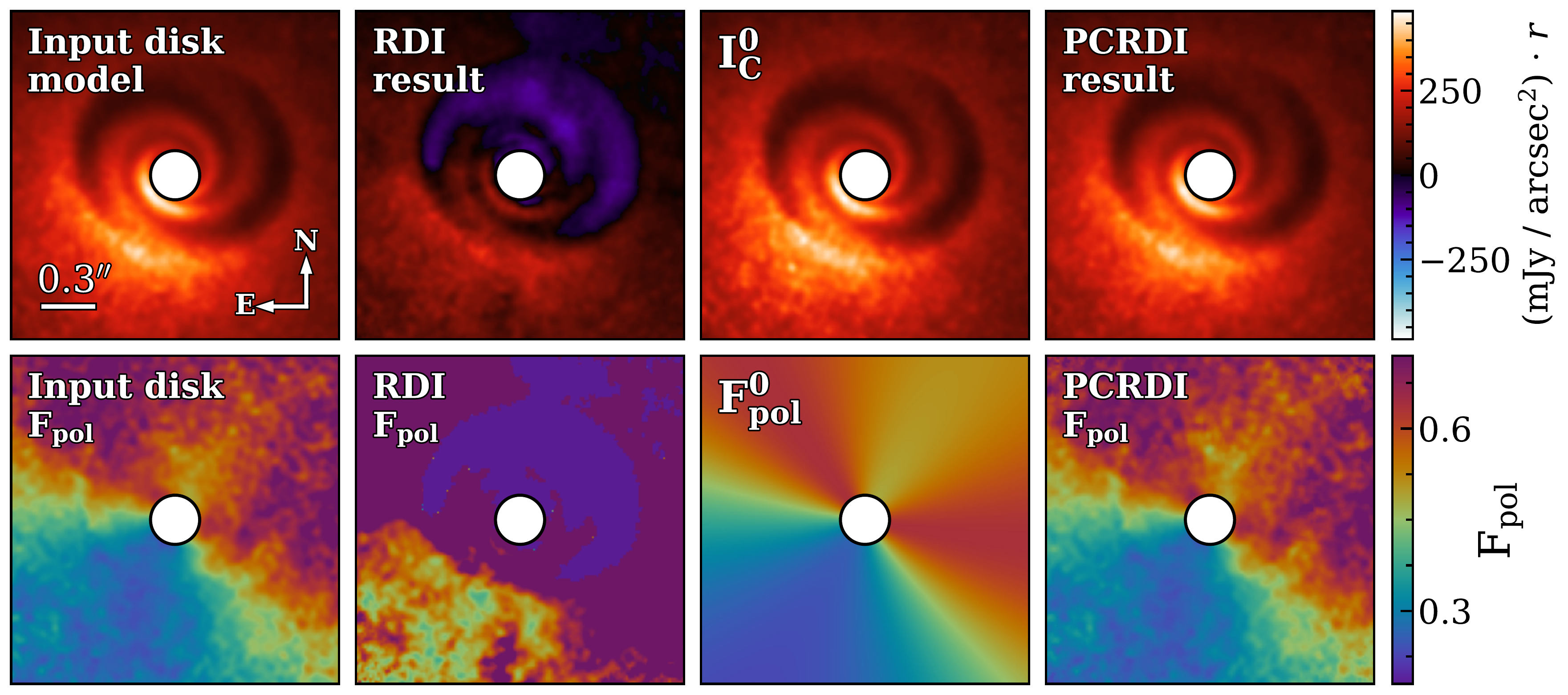}
    \caption{Wavelength-averaged results for the simulated data described in Section \ref{sec:sims}. \textbf{Top row}, from left to right: the \ac{psf}-convolved input disk model image (the desired result), the heavily-attenuated final image using RDI \ac{psf}-subtraction, the optimized PI-based estimate of the disk signal used by PCRDI, and the final image using PCRDI \ac{psf}-subtraction (Section \ref{sec:pcrdi}). While the disk signal estimate used by PCRDI is imperfect, PCRDI is still able to recover an extremely high-fidelity image of the disk. \textbf{Bottom row}: the corresponding fractional polarization map for each result in the top row. The RDI result is non-physical over most of the field of view ($F_{pol} > 1$ or $F_{pol} < 0$). Though the ``smooth surface'' $F_{pol}$ estimate (center-right) used to derive $I_C^0$ is much less detailed than the $F_{pol}$ for the input disk model (left), the PCRDI result (right) still recovers the very fine polarization gradients seen in the input model.
    \label{fig:pcrdi_sim_results}
    }
\end{figure*}

\begin{figure}
    \centering
    \includegraphics[width=0.47\textwidth]{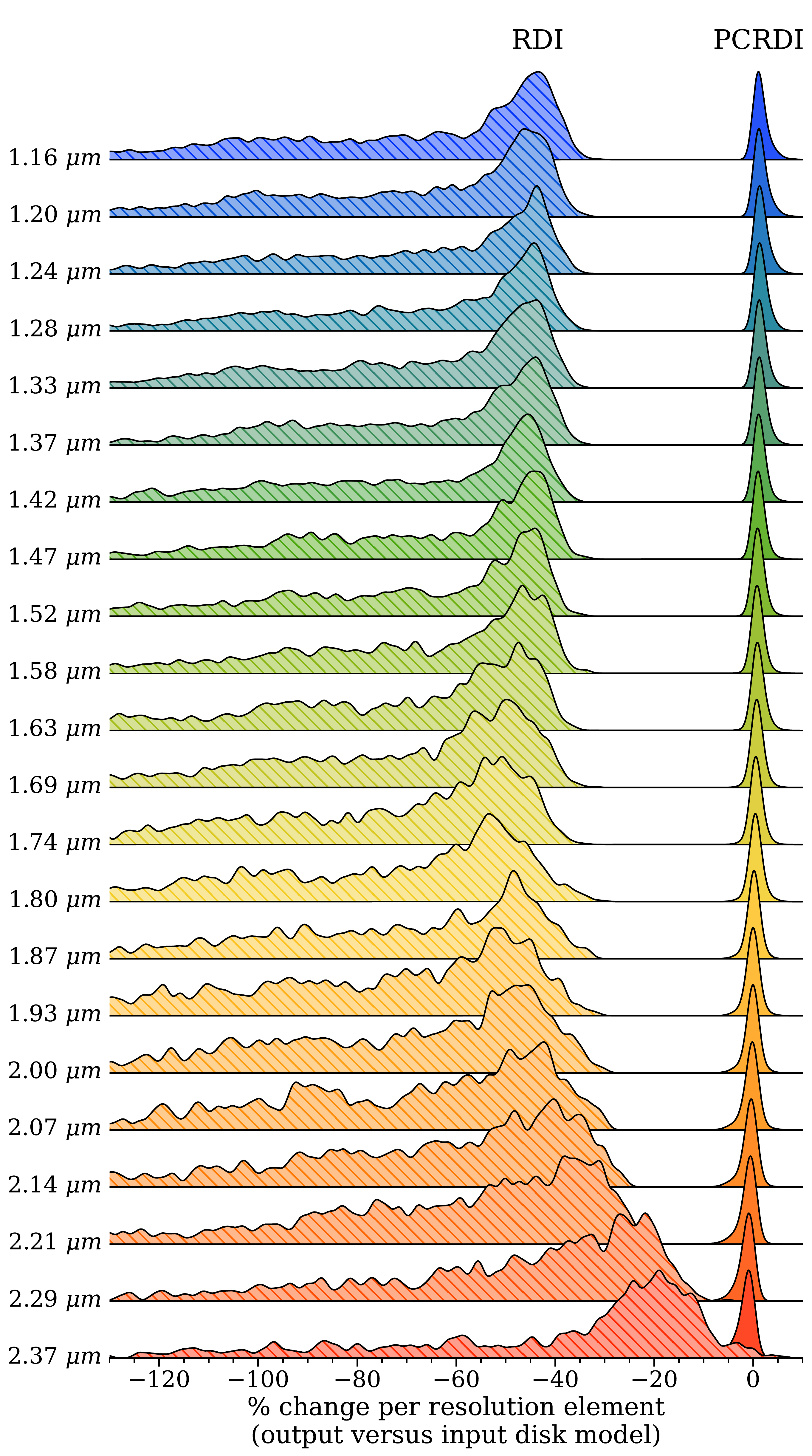}
    \caption{For the simulated data of Section \ref{sec:sims}: distributions (as kernel density estimates) of percent change in disk signal (relative to the `ground truth' input disk model) due to attenuation for each CHARIS wavelength channel. For this purpose, all results are binned to the resolution of that wavelength channel prior to the calculation. Hatched and solid regions correspond to results for the standard and PI-constrained RDI reductions respectively. RDI results show significant and highly variable attenuation (within a single wavelength channel and between wavelength channels). Meanwhile, PCRDI results are uniformly within roughly a percentage point of the true values. 
    \label{fig:pcrdi_sim_kde}
    }
\end{figure}

The results of PCRDI optimization are shown in Figure \ref{fig:pcrdi_sim_results} alongside those of the standard (unconstrained) \ac{rdi} procedure. While the best-fit parameter set produces a fractional polarization model that is much less detailed than that of the (PSF-convolved) input model, it nevertheless enables an extremely high fidelity fractional polarization map to be computed using the disk PI and the PCRDI total intensity result. We note that small differences between the input disk images and the PCRDI total intensity images remain. However, binning the images to resolution and computing the percent difference per resolution element shows differences generally within $\sim 1\%$ of zero (see Fig. \ref{fig:pcrdi_sim_kde})\footnote{Note: to provide attenuation estimates which are not significantly dependent on the quality/stability of the \ac{ao} correction throughout the observations, this assessment and the distributions plotted in Fig. \ref{fig:pcrdi_sim_kde} exclude any residual starlight -- such that oversubtraction is the sole source of any change from ground truth.}.

{When using attenuated disk images for analysis of disk surface brightness and color, forward modeling is often used to derive approximate attenuation corrections \citep[e.g.,][]{Goebel2018, Betti2022}. To provide a more direct comparison of constrained RDI with such techniques, we derive and apply attenuation corrections for the conventional \ac{rdi} reduction in a similar manner (see Appendix \ref{app:rdi_corrections} for details) using two different disk models. For the first model, we use the ``ground truth" disk model contained in the simulated data --- referred to hereafter as the ``ideal" case. This should be regarded as an absolute upper limit on the accuracy of corrections derived this way; for similarly complicated disks, modeling results do not typically approach this accuracy, and often do not attempt to reproduce non-axisymmetric structures at all \citep[e.g.,][]{Lomax2016, Betti2022}. The second model instead emulates a more realistic modeling result, consisting of a simple two ring disk model without any spiral structures but which generally reproduces the bulk of the disk's shape and brightness --- referred to hereafter as the ``realistic" case. Both models are visualized in Figure \ref{fig:rdi_vs_pcrdi_fm} of Appendix \ref{app:pcrdi_fwdmod} (as ``Model 1" and ``Model 2", respectively), with the forward modeled results for \ac{rdi} presented in the top row.}

\begin{figure*}
    \centering
    \includegraphics[width=0.98\textwidth]{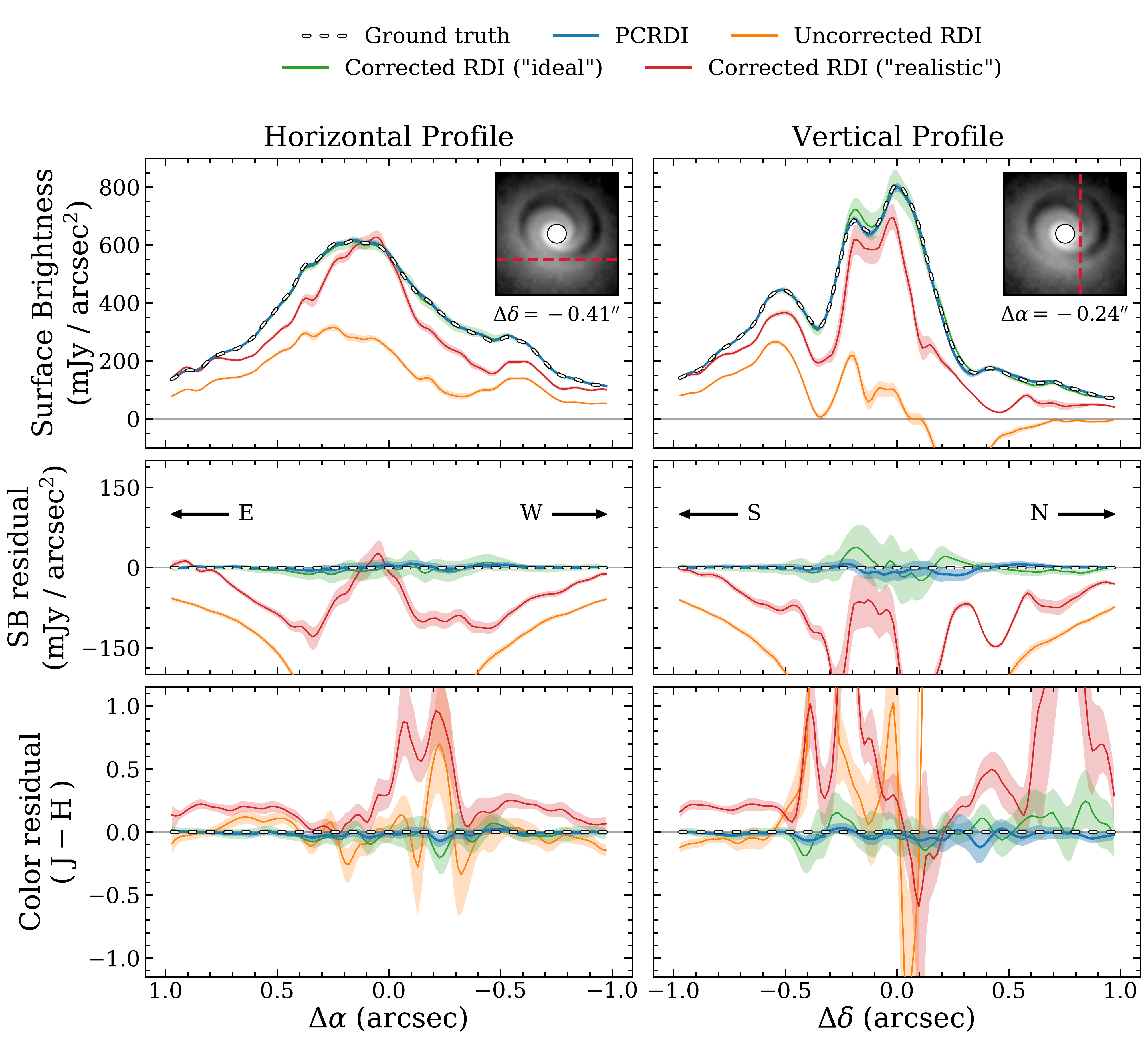}
    \caption{Surface brightness (SB) and color measurements for the results of Section \ref{sec:pcrdi_sims} using an aperture with a 2 pixel radius (diameter $\sim 0\farcs065$, slightly larger than the FWHM of the longest CHARIS wavelength channel: $0\farcs062$). Shaded regions indicate approximate 3$\sigma$ confidence intervals following the procedure of Appendix \ref{app:sb_uncertainty}. \textbf{Top row}: Broadband SB along a horizontal (left) and vertical (right) profile (the dashed red line in the inset images). \textbf{Center row}: The difference between each reduction and the ground truth profile for the measurements in the top row. \textbf{Bottom row}: Offset of $J-H$ disk color from that of the ground truth model for the same products and positions. Gaps in the data (particularly for the vertical profile of the uncorrected RDI reduction) correspond to undefined color measurements resulting when one of the two filters measured a negative surface brightness.
    \label{fig:sb_comp_pcrdi}}
\end{figure*}

{Figure \ref{fig:sb_comp_pcrdi} compares the ground truth surface brightness (SB) and color profiles with those measured from the corrected RDI results, as well as the uncorrected RDI and the PCRDI results. For this purpose, we bin each of the final image cubes along the wavelength axis to approximate \ac{nir} J-, H-, and K-bands. Additionally, we take the average over the entire CHARIS wavelength range to produce a single image which we refer to simply as ``broadband". Overall, PCRDI significantly outperforms model-corrected RDI --- even in the unlikely scenario in which modeling \emph{exactly} recovers the underlying disk. The inaccuracy that remains even in this ``ideal" correction scenario is the result of a key assumption made by the forward modeling correction strategy: that only attenuated disk signal remains in the image being corrected. In fact, there is also residual starlight and noise --- both of which are inflated when the disk signal is corrected, leading to the inaccuracy measured here. For the $J-H$ color profiles, comparing measurements for PCRDI with those of RDI with ``ideal" corrections: PCRDI is, on average, closer to the ground truth by a factor of $\sim 2$ for the horizontal profile, and by a factor of $\sim 4$ for the vertical profile (with uncertainties which are smaller by factors of $\sim 4$ and $\sim 5$ respectively). Forward modeling the PCRDI \ac{css} estimate, $I_C^0$, to make corrections for RDI yields comparable accuracy to the ``ideal" corrections in the inner disk, which become somewhat worse beyond the region in which the estimate was optimized ($\sim0\farcs45$). These results demonstrate that, in any regions where starlight and/or noise are non-negligible compared to the residual \ac{css}, application of corrections to a conventional RDI result will yield less accurate measurements than constrained RDI. In other words, while correcting RDI this way cannot improve the ratio of \ac{css} to a) noise or b) residual starlight at a given location, use of constrained RDI can.}

A procedure for PCRDI forward modeling is provided in Appendix \ref{app:pcrdi_fwdmod}. However, for any typical application, analysis can be performed on PCRDI products without the need for extensive forward modeling. 

\subsection{MCRDI Throughput}
To demonstrate the MCRDI technique, we carried out simplified simulations of JWST NIRCam F335M observations of the debris disk system HD 10647\footnote{Based on upcoming JWST Cycle 1 \ac{gto} observations of this system (Program 1183).}. The NIRCam \ac{psf} was simulated using WebbPSF\footnote{\url{https://github.com/spacetelescope/webbpsf}} \citep{Perrin2014} while the debris disk was simulated using DiskDyn \citep{Gaspar2020}. The NIRCam PSF simulations include Gaussian telescope jitter ($\sigma = 7$ mas) and pointing error ($\sigma = 4$ mas per axis), and include two roll angles for the target ($\Delta\rm{PA} = 10\degr$) and a 5-point dither of the reference star, but neglect other factors such as noise and thermal effects. For simplicity, the \ac{psf} reference star's spectrum is assumed to be identical to that of the target --- corresponding to a perfect color and flux match. These simulations should provide a reasonable comparison of the relative performance of the \ac{psf} subtraction techniques, but are not representative of the absolute quality of JWST/NIRCam products. 

To generate the disk models used as constraints in MCRDI, we utilized the GRaTeR debris disk modeling code \citep{Augereau1999}, parameterized as described in \citet{Lawson2021b} (with the addition of a parameter for the disk's brightenss) {and using a Hong-like scattering phase function (\citealt{Hong1985}; in contrast to the  DiskDyn model, in which scattering properties are simulated for a disk of astronomical silicate dust for particles evenly spaced in mass/size log space between 0.1$\micron$ and 1cm)}. Using a different tool than the one used to generate the ``true'' input disk model is intended to better emulate real observations, where a model is unlikely to perfectly describe the data. For optimization of the disk signal estimate, we again used the Levenberg–Marquardt algorithm. Given the differences in parameterization between the two tools (and the fact that DiskDyn is a dynamical code, such that the values of some parameters for the final evolved model differ from their initializations), the optimal GRaTeR parameter values are not known \textit{a priori} --- with the exception of inclination and position angle (PA). For the MCRDI optimization, we initialized inclination and PA with values offset from the true DiskDyn values by the $1\sigma$ uncertainties for these parameters reported in \citet{Lovell2021} ($1\degr$ in both cases). The remaining parameters are initialized using rough `by-eye' estimates of their values from an unconstrained (i.e., conventional \ac{rdi}) reduction of the data. The proximity of the initial values to the optimal values may affect the number of function evaluations required to reach a solution but will not affect the quality of the MCRDI result (unless the parameter space has multiple optima, in which case a global optimization algorithm would be warranted instead). For general use, values from disk modeling performed in prior studies could also be adopted.

\begin{figure*}
    \centering
    \includegraphics[width=0.99\textwidth]{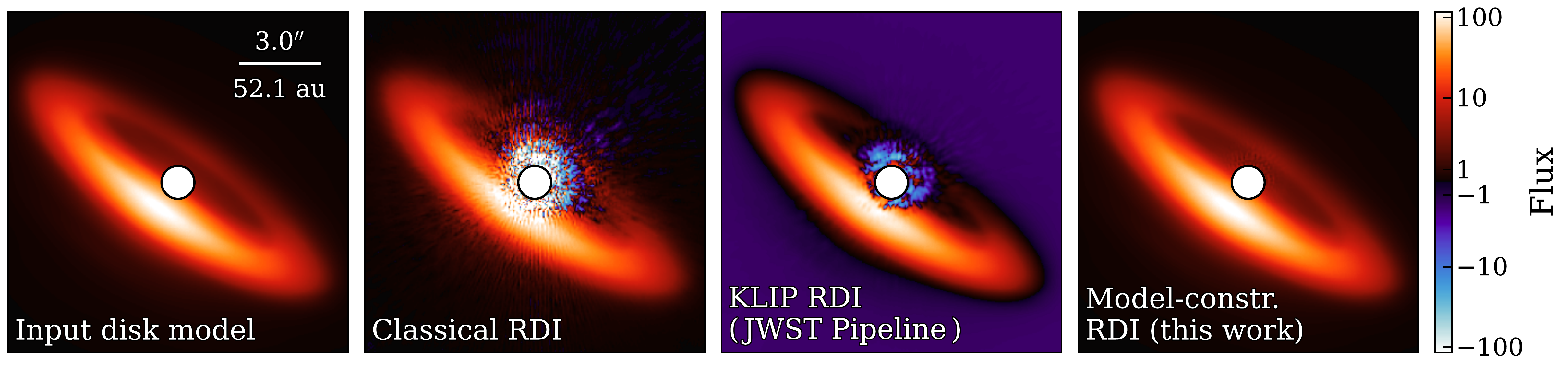}
    \caption{Simulated results for JWST NIRCam observations of the debris disk system HD 10647. The leftmost panel shows the input disk model (the ideal result), with the subsequent panels showing the result using the indicated subtraction technique. Model constrained RDI (MCRDI) meets or exceeds the fidelity of classical RDI at large separations, while performing substantially better than both classical and KLIP subtraction at small separations (see also: Fig. \ref{fig:jwst_sim_sb}).
    \label{fig:jwst_sim_ims}}
\end{figure*}

\begin{figure*}
    \centering
    \includegraphics[width=0.85\textwidth]{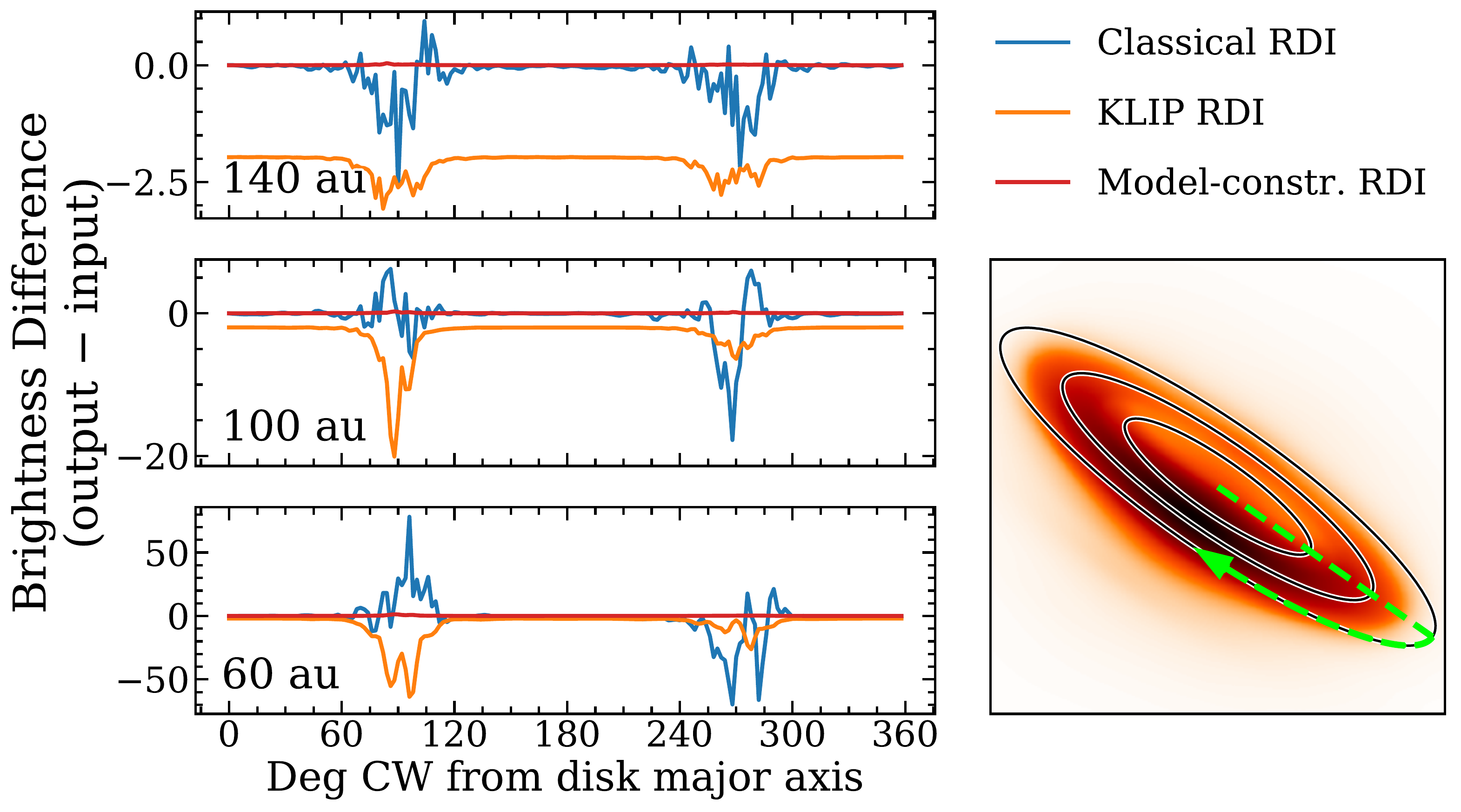}
    \caption{For each reduction of Fig. \ref{fig:jwst_sim_ims}: the difference between the output and input brightness of the disk measured in apertures with diameter equal to the PSF full-width at half maximum. The lower-right panel provides a visualization of the three radial separations measured (black ellipses), and also shows the origin and direction of the measurements (green arrow). At $r=140$ au and $r=100$ au, classical RDI and model constrained RDI (MCRDI) both reasonably reproduce the majority of the ``ground truth" brightness profile. MCRDI provides substantially improved measurements throughout the inner region ($r=60$ au), corresponding to the inner asteroid belt for this system, and the most likely location of any yet-unseen companions.
    \label{fig:jwst_sim_sb}}
\end{figure*}

A comparison of the results using classical RDI\footnote{For classical RDI, we use the median of the reference sequence as the \ac{psf} model. More so than KLIP or MCRDI, the efficacy of classical RDI depends on factors which are difficult to generalize here. For example, the amount of residual starlight is dictated in part by the randomly determined positions of the target and reference stars behind the coronagraph (the pointing offset). For the small number of images used, a given sequence could plausibly perform noticeably better or worse than the result here. However, the qualitative relationship will remain --- i.e., that classical RDI provides high \ac{css} throughput, but worse starlight suppression at smaller separations.}, KLIP RDI\footnote{As implemented in the official JWST pipeline and retaining $N=10$ KLIP modes; the quality of the result does not improve noticeably for other choices ($N=1-15$). Using the unconstrained RDI procedure of Section \ref{sec:pcrdi_sims} instead yields comparable results.}, and model constrained RDI are shown in Figures \ref{fig:jwst_sim_ims} and \ref{fig:jwst_sim_sb}. The optimization region used for both KLIP and MCRDI excludes only the region within 11 pixels of the star (0\farcs69, the approximate inner working angle). These results show that, at moderate to large separations, classical and constrained reference subtraction both reasonably reproduce the ``ground truth" brightness profile -- with MCRDI performing somewhat better and KLIP performing significantly worse. At small separations, MCRDI performs substantially better than the other techniques.

{In this case, we do not compare MCRDI with model-corrected RDI results as we did in the previous section. Since our simulated data do not include realistic noise, this would not provide a reasonable estimate of the quality that can be achieved with such corrections.}

\section{Application of PCRDI to IFS Observations of AB Aurigae}\label{sec:abaur}

AB Aurigae is a pre-main sequence star (1--3 Myr, \citealt{Kenyon2008taurusauriga}; $d=156$ pc, \citealt{Gaia2016, Gaia2021}) hosting a highly structured protoplanetary disk that presents with spiral arms at both large scales \citep[hundreds of au;][]{Grady1999, Fukugawa2004} and small scales scales \citep[tens of au;][]{Hashimoto2011, Boccaletti2020} in optical and \ac{nir} imagery, and with a large, depleted inner cavity in sub-millimeter \citep{vanderMarel2021}.

As a further demonstration of the PCRDI technique, we utilize \ac{scexao}/\ac{charis} observations of AB Aur in total intensity ($t_{int} \approx 116$ minutes) and polarized intensity ($t_{int} \approx 74$ minutes) -- originally reported in \citet{Currie2022} and \citet{Lawson2021spie} respectively. These data were taken over two consecutive nights in October 2020, with both polarized and total intensity sequences using \ac{charis} in low-res broadband mode ($R \sim 19$, $1.15 - 2.39$ $\mu m$ -- producing 22 wavelength channels per exposure) and utilizing a 113 mas Lyot coronagraph. \citet{Currie2022} also observed a reference star (HD 31233) before and after the AB Aur total intensity sequence ($t_{int} \approx 21$ minutes), which we also use here to enable \ac{rdi}. See \citet{Currie2022} for a description of preprocessing (sky subtraction, image registration, etc.) used for the total intensity data, and \citet{Lawson2021spie} for a description of the \ac{pdi} data reduction for the \ac{pi} data.

As with the synthetic data, \ac{rdi} \ac{psf}-subtraction is performed by constructing the \ac{psf} model for each target frame using a linear combination of the reference frames and comparing the target and reference data within an annular optimization region spanning $r= 10-25$ pixels ($\sim 0\farcs16-0\farcs41$)\footnote{{The conventional RDI reduction is different than (but comparable to) the one in \citet{Currie2022}, which used KLIP with larger optimization regions instead.}}. For this data, the relatively narrow optimization region is necessary to accommodate the following:
\begin{enumerate}
    \item \Ac{charis} \ac{pdi} data has a rectangular $\sim 1\arcsec \times 2\arcsec$ \ac{fov} (compared to the square $2\arcsec\times2\arcsec$ \ac{fov} for the classical mode) and is collected in pupil-tracking mode. The final sequence-combined \ac{pi} image lacks coverage of two wedge-shaped regions extending from $r\sim0\farcs5$ \citep[see][]{Lawson2021spie}. Limiting the optimization region to separations with full azimuthal coverage substantially simplifies and accelerates \ac{psf} model optimization.
    \item \citet{Hashimoto2011} report a misalignment between AB Aur's inner disk ($\sim 0\farcs21 - 0\farcs43$) and outer disk ($\sim 0\farcs63- 0\farcs84$), which would inhibit our smooth-scattering surface estimate for an optimization region extending to the outer disk.
    \item AB Aur shows a substantial total intensity enhancement at $r\sim0\farcs6$ (nearly directly south of the star) which is not reproduced by scaling the \ac{pi} image with reasonable fractional polarization models. This feature is analyzed further in \citet{Currie2022}, where it is determined to be an embedded protoplanet. Avoiding this region by optimizing at smaller separations enables a higher fidelity estimate of the total intensity.
\end{enumerate}

\begin{figure*}
    \centering
    \includegraphics[width=0.85\textwidth]{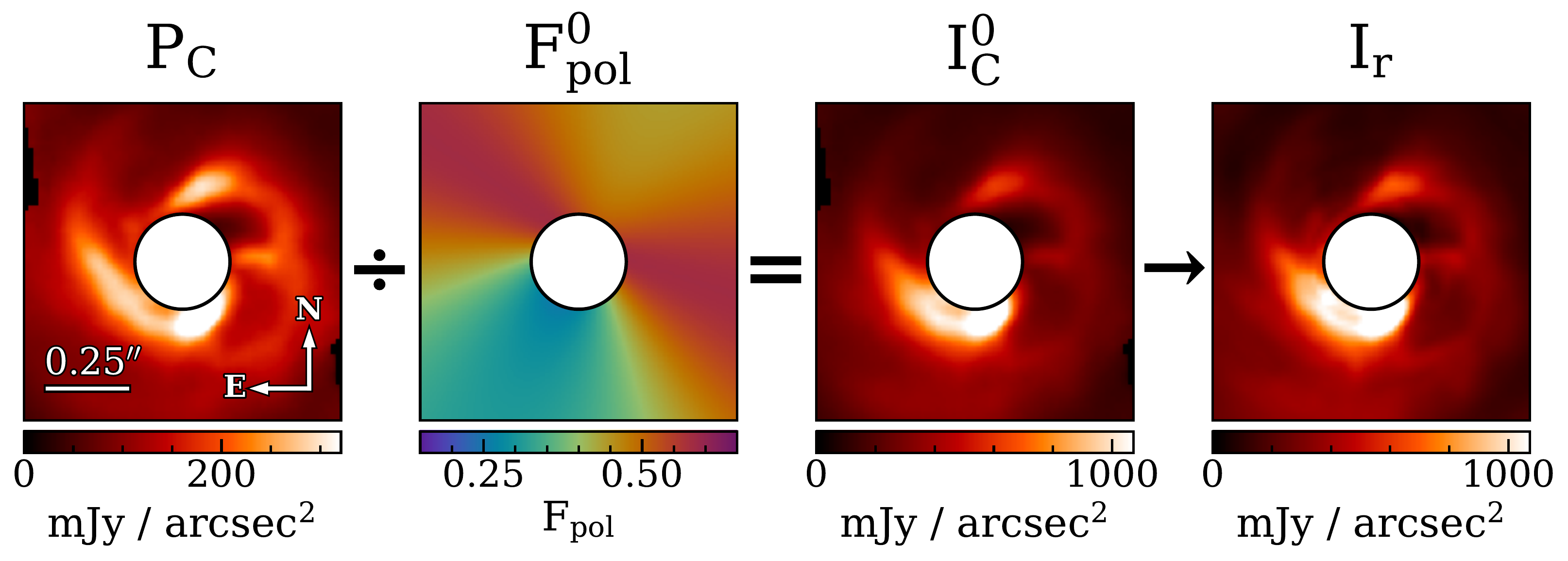}
    \caption{The components of the PCRDI procedure (Section \ref{sec:pcrdi}) as utilized for AB Aur. Left: the polarized intensity disk image ($P_C$); center-left: the best-fit fractional polarization model by which $P_C$ is divided to estimate the total intensity of the disk; center-right: the total intensity estimate of the disk ($I_C^0$); right: and the final sequence-combined residual image from PCRDI.
    \label{fig:abaur_pcrdi_procedure}}
\end{figure*}

\begin{figure*}
    \centering
    \includegraphics[width=0.99\textwidth]{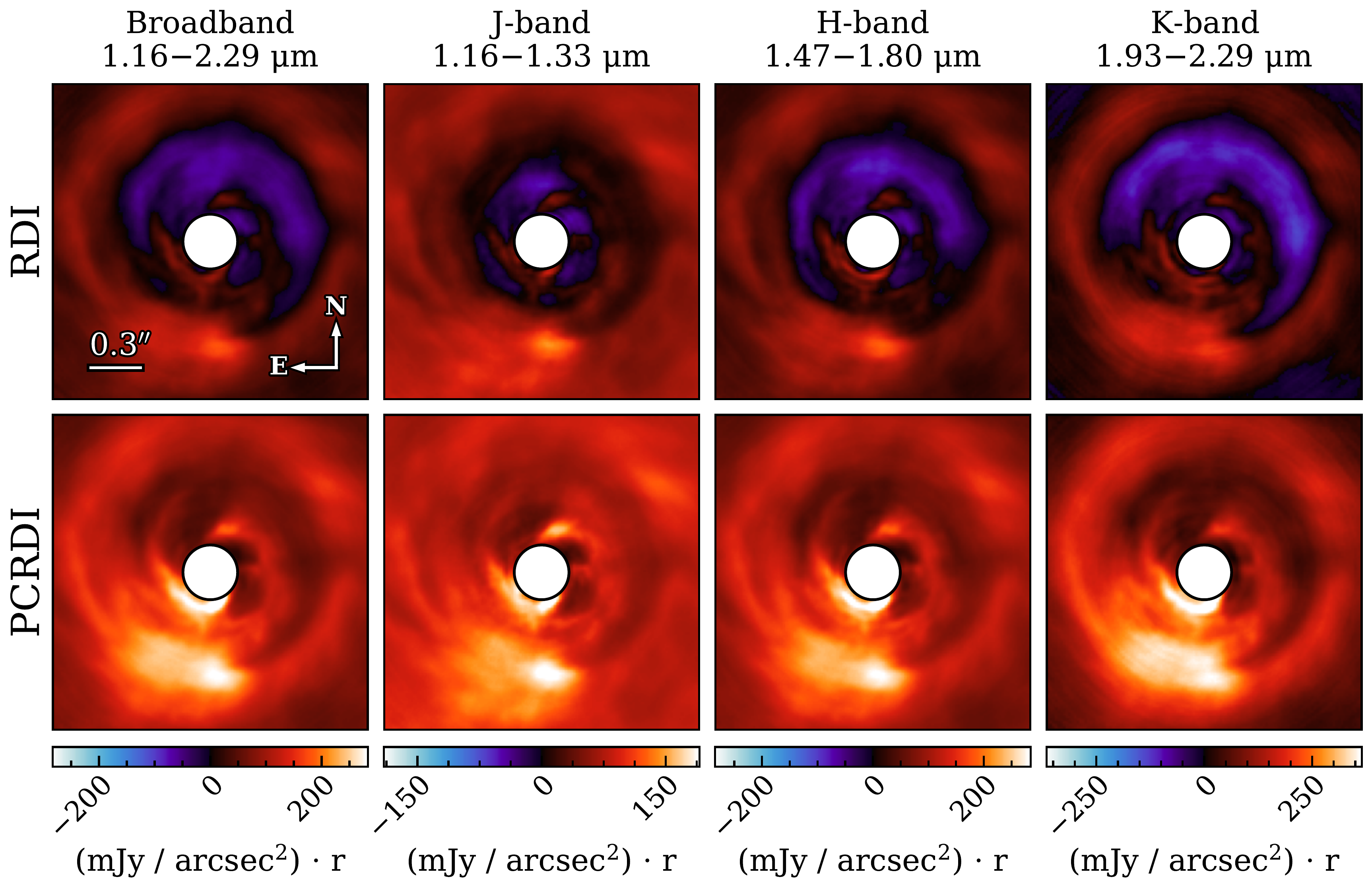}
    \caption{A comparison of PSF-subtracted results for SCExAO/CHARIS observations of AB Aur (see Section \ref{sec:abaur}) using a conventional RDI procedure (top row) and PI-constrained RDI (PCRDI; bottom row). The PCRDI results (see Sections \ref{sec:pcrdi}, \ref{sec:abaur}) use the same PSF-subtraction algorithm and algorithm settings as the conventional RDI result, but adopt an optimized PI-based constraint for the circumstellar signal to mitigate oversubtraction. All images are multiplied by the stellocentric separation in units of arcseconds and assuming an orientation and geometry appropriate for the inner disk. The columns correspond to different binnings over the wavelength axis of the final CHARIS IFS image cube; from left to right: JHK broadband ($1.16-2.29$ $\micron$), $J$-band (channels $1-5$, $1.16-1.33$ $\micron$), $H$-band (channels $8-14$, $1.47-1.80$ $\micron$) and $K$-band (channels $16-21$, $1.93-2.29$ $\micron$).
    \label{fig:abaur_pcrdi}}
\end{figure*}

\begin{figure}
    \centering
    \includegraphics[width=0.47\textwidth]{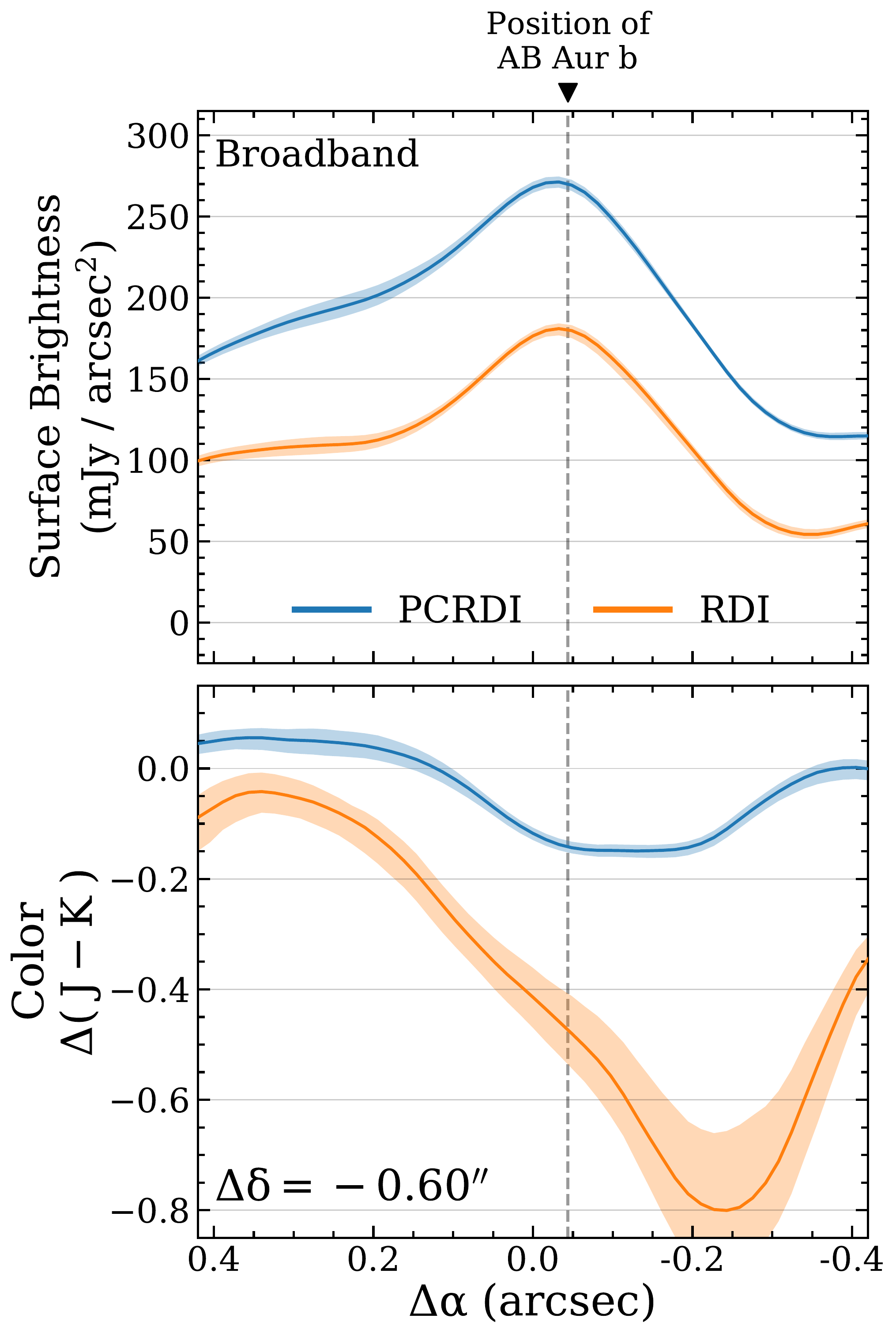}
    \caption{A comparison of surface brightness and color profiles across the position (east-west) of the confirmed protoplanet AB Aur b for the conventional RDI (orange) and PCRDI (blue) reductions; see Section \ref{sec:abaur}. The shaded regions are approximate 3$\sigma$ confidence intervals (see Appendix \ref{app:sb_uncertainty}). \textbf{Top}: Broadband surface brightness measured using an aperture with a radius of 4 pixels (0\farcs065; for consistency with \citealt{Currie2022}). The two results show similar overall shapes, but with the standard RDI result being significantly fainter due to oversubtraction. \textbf{Bottom}: As above, but for the $J-K$ color of the disk in magnitudes (with the color of the parent star subtracted). The PCRDI result shows a bluer color at the position of AB Aur b relative to the disk on either side. While the RDI result also shows a blue dip, it is a) of much lower significance, and b) manifests as a smoother color transition that is not clearly distinct around the location of AB Aur b.
    \label{fig:abaur_sb_and_color}}
\end{figure}

A visualization of the polarized intensity, the final optimized fractional polarization estimate ($F_{pol}^0$), the corresponding PI-based \ac{css} estimate ($I_{C}^0$), and the final PCRDI result ($I_r$) is provided in Figure \ref{fig:abaur_pcrdi_procedure}. Figure \ref{fig:abaur_pcrdi} compares the PCRDI results with the conventional RDI results in an array of wavelength ranges -- where, as for the simulated data (Section \ref{sec:sims}), PCRDI provides an improvement in \ac{css} throughput that is immediately apparent. Notably, in the PCRDI imagery it is evident that the morphological feature $\sim 0\farcs6$ south of the image center \citep[AB Aur b;][]{Currie2022} has a distinct (bluer) spectral energy distribution relative to the nearby disk -- appearing at much higher contrast to the disk in $J$-band than in $K$-band.
Similar conclusions are more difficult to draw from the conventional RDI imagery as a result of the substantial spatially- and spectrally- variable oversubtraction that occurred during PSF subtraction. This effect is visualized in Figure \ref{fig:abaur_sb_and_color} using measurements of surface brightness and color across the position of AB Aur b. 

To further validate this result (or PCRDI results in general), an additional test using forward-modeling for the conventional \ac{rdi} reduction is summarized in Appendix \ref{app:fwdmod}. 

\section{Broader Applications}\label{sec:broader_apps}

Both variations of constrained \ac{rdi} presented here -- PCRDI and MCRDI -- can be applied to substantially improve the products of many disk studies. Moreover, constrained \ac{rdi} can be combined with many existing algorithms for construction of the stellar \ac{psf} model. As such, this strategy can be implemented in existing pipelines with very little alteration necessary; the only change needed in most cases is the ability to model the stellar \ac{psf} using a different (\ac{css}-estimate-subtracted) target sequence than the one from which the \ac{psf} model is ultimately subtracted. For reference, the run-times for the reductions carried out in this work are provided in Appendix \ref{app:runtimes}.

Though we have demonstrated the use of PCRDI with near-contemporaneous \ac{pi} data from the same instrument, the same principles can be applied using non-contemporaneous \ac{pi} data from an entirely different instrument. For example, when using the $H$-band VLT/SPHERE \ac{pi} imagery of AB Aur reported in \citet{Boccaletti2020} to carry out PCRDI on the \ac{charis} \ac{rdi} data, the result is nearly identical. Considering the prolific extent of prior extreme adaptive optics \ac{pi} disk surveys \citep[e.g., with GPI and SPHERE; ][]{Esposito2020, Garufi2022, Rich2022}, this feature provides PCRDI with broad utility for nearly any group conducting high-contrast imaging studies of disks. This provides a compelling application for data from instruments without polarimetric imagers as well -- including upcoming \ac{jwst} data -- for which ground-based \ac{pi} imagery can be used to improve \ac{psf}-subtraction. 

Similarly, MCRDI can be applied for any system whose disk can be \emph{superficially} reproduced with a synthetic model. In other words, the utilized model need not be fully physically motivated. For example, a simple scattered light disk model (physically appropriate for a debris disk) could be used to suppress oversubtraction for observations of a transitional disk (whose appropriate physical model would normally require a more involved radiative transfer framework). In such a scenario, the procedure for constrained RDI forward modeling described in Appendix \ref{app:pcrdi_fwdmod} could be adopted. Here, faster scattered light models could be used to optimize the MCRDI result, with forward modeling of more physically motivated models (perhaps using the MCRDI model parameters to initialize the procedure) then being used as the basis for interpretation of the imagery.

Additionally, for \ac{rdi} data for which a full forward modeling procedure has been performed, MCRDI can be applied retroactively using the model identified from forward modeling to improve the RDI results. {In fact, the residuals when evaluating an MCRDI CSS estimate (Equation \ref{eq:opt_pcrdi_obj}) are identical to the forward modeled residuals (processed data$-$model) when adopting the MCRDI constraint as the model --- so long as $\mathcal{M}(I, \mathcal{R})$ is linear in $I$ (e.g., KLIP and LOCI). I.e., if $\mathcal{M}(I_1 - I_2, \mathcal{R}) = \mathcal{M}(I_1, \mathcal{R}) - \mathcal{M}(I_2, \mathcal{R})$, then (starting from the right-hand side of Equation \ref{eq:opt_pcrdi_obj}):}
\begin{equation}
    \begin{split}
    (I-I_C^0) - \mathcal{M}(I - I_C^0, \mathcal{R}) \\
    = (I-I_C^0) - (\mathcal{M}(I, \mathcal{R}) - \mathcal{M}(I_C^0, \mathcal{R})) \\
    = (I-\mathcal{M}(I, \mathcal{R})) - (I_C^0 - \mathcal{M}(I_C^0, \mathcal{R})),
    \end{split}
\end{equation}
{which is the standard equation for forward modeled residuals for model $I_C^0$ with a conventional RDI procedure. In other words, model parameters from forward modeling optimization should be identical to those from MCRDI optimization. Following from the analysis and discussion comparing constrained RDI and model-corrected RDI in Section \ref{sec:pcrdi_sims}: when a strong \ac{css} model has been identified, it appears that an MCRDI reduction using that model as the constraint should improve measurement accuracy over application of model-based corrections to the conventional RDI product.}

{Unlike examples presented for \emph{REXPACO} \citep{Flasseur2021} and \emph{MAYONNAISE} \citep{Pairet2021}, the constrained RDI examples presented here make no effort to directly disentangle distinct sources of \ac{css} (e.g., separating planet from disk signal). In application to AB Aur, we have simply avoided the vicinity of AB Aur b while constructing the PSF model and then relied on spectrophotometric analysis (afforded by the \ac{ifs} data) to distinguish AB Aur b from the disk. In principle, however, a point-source could be added to the \ac{css} estimate used by constrained RDI --- optimizing its location and brightness alongside any parameters governing the estimate of the disk. Alternatively, constrained RDI could conceivably be combined with one of these source-separation tools --- e.g., by using the constrained \ac{rdi} \ac{psf} model in place of the \ac{psf} model from the ``fixed-point algorithm" in \emph{MAYONNAISE}.}

\section{Considerations and Limitations}\label{sec:limitations}

{As noted, our PI-based estimate of the total intensity of \ac{css} assumes a simple ``bell-shaped" fractional polarization curve which peaks at a scattering angle of $90\degr$ (Eq. \ref{eq:rayleigh_pol}). This prescription is generally appropriate for \ac{nir} observations of disks --- irrespective of the relation between the dominant dust size and the observing wavelength \citep[with larger grains tending to produce diminished fractional polarization, but still maintaining a bell shape and center about $90 \degr$;][]{Benisty2022}. However, even if the true fractional polarization peaks at a different scattering angle or manifests with a skewed shape \citep[e.g., the C01 model of][]{Takami2013}, the utilized model may still be sufficient for a strong PCRDI solution. By tuning the parameters governing the assumed scattering surface (away from those of the ``true" surface), the utilized curve can emulate fractional polarization maps for other polarization curves over a region of interest. For example, an inclined disk with a particular surface flare ($c$) and a polarization curve with a peak skewed from $90\degr$ toward higher or lower scattering angles could be reproduced by increasing or decreasing (respectively) the assumed flare of the scattering surface without altering the assumed polarization curve.} By allowing our optimization to explore a wide range of scattering surfaces, the polarization model of Eq. \ref{eq:rayleigh_pol} can provide strong solutions even for disks with polarization that is not truly well-described this way.

Nevertheless, care should be taken when applying this model to highly inclined and/or highly flared disks \citep[where a wider range of scattering angles may be probed; e.g., IM Lup,][]{Avenhaus2018} which also have unconstrained or otherwise non-bell-shaped polarization curves. In such a scenario, results may be improved by utilizing a more appropriate polarization curve in place of Eq. \ref{eq:rayleigh_pol}. {This may be particularly relevant for observations at optical wavelengths, where fractional polarization curves can take more complex ``rippled" shapes for some compositions \citep[e.g., for the case of amorphous carbons in][]{Tazaki2022}.} 

Given these considerations, we would also recommend against interpreting the best-fit scattering surface parameters from optimization of PCRDI (inclination, etc.) as the ``true" values unless the disk's fractional polarization curve is well-constrained (and an appropriate model utilized in fitting) and its surface is consistent with the assumed scattering surface model. Moreover, unless both the scattering surface \emph{and} the fractional polarization curve for a given disk are well constrained, utilizing the optimization method outlined in Section \ref{sec:optimization}, rather than adopting literature parameters, will typically be more effective for eliminating oversubtraction. {In cases where there may be differences in the flux calibration between the polarized and total intensity, the peak fractional polarization can be optimized to offset any impact this might otherwise have on PCRDI throughput.}

The strictest requirement for use of PCRDI is simply that the disk be detected in \ac{pi} within the optimization region to be used for total intensity \ac{psf} subtraction. Specific limits with respect to disk morphology and orientation will depend significantly on the details of the system and the data. For example, since PCRDI scales the \ac{pi} image, any noise will be significantly inflated in regions where the assumed fractional polarization is low (e.g., scattering angles far from $90\degr$ for a bell-shaped polarization curve). Though this noise will not directly enter the final PCRDI result\footnote{As noted in Section \ref{sec:constrained_rdi}, the \ac{css} estimate is never directly used in the final result.}, it may complicate optimization for disks with weak \ac{pi} signal-to-noise ratio and low fractional polarization in the optimization region. We relegate a more complete investigation of any geometric and morphological limits related to PCRDI to a later study.

In the case of MCRDI, the sole prerequisite is that the underlying circumstellar signal can be superficially reproduced with synthetic disk models; there are no additional considerations related to the morphology or orientation of the disk beyond those relevant for typical forward modeling procedures. As for PCRDI, we hazard against directly adopting the disk parameters that result from MCRDI optimization unless a fully physically-motivated model has been used.

Finally, we note that our examples and analysis utilize small, near-contemporaneous sequences of reference images and target data with significant field rotation. Additional considerations may be motivated for applications to dissimilar data. {When using extremely large \ac{psf} libraries spanning many nights \citep[e.g., such as those used for ALICE;][]{Choquet2014, Hagan2018}, it may be possible to manifest features of the \ac{css} estimate which are not truly present in the data from noise or artifacts in the reference data. In practice, this can be avoided by limiting the size of the \ac{psf} library in some way (e.g., by using KLIP and retaining only part of the KLIP decomposition of the reference library) and adopting frame selection practices similar to those developed for ALICE \citep[e.g., to remove reference images containing \ac{css} or artifacts that might mimic circumstellar structure;][]{Choquet2014, Hagan2018}. Likewise, use of \ac{nmf} \citep{Ren2018} with constrained RDI (in place of LOCI or KLIP) may help to further insulate against false-positive \ac{css} features for such data; since an erroneous feature in the \ac{css} estimate would manifest as a negative imprint in the \ac{css}-subtracted data for which the starlight model is constructed, reconstruction of such a feature from noise in reference images should be much less likely if negative reference coefficients are not permitted. Ultimately, the best cases for application of constrained RDI are those showing significant but badly oversubtracted \ac{css} when using conventional PSF-subtraction techniques. If the conventional \ac{rdi} result is effectively nulled, then any significant \ac{css} must apparently be well-matched by the PSF model with the permitted model freedom, and so any features appearing following application of constrained RDI may be unreliable.}
{In such scenarios, the forward modeling procedure of Appendix \ref{app:pcrdi_fwdmod} could also be used to test against false-positive features resulting this way --- e.g., by forward modeling a model lacking the feature and checking if said feature manifests in the result anyway due to the constraint and the freedom of the PSF model. We emphasize, however, that this possibility is purely speculative; this has not occurred for any of our testing thus far.}

\section{Conclusions}
Herein, we have presented ``constrained RDI", a new class of \ac{rdi} \ac{psf} subtraction techniques well-suited for circumstellar disk targets. Using variations with both \ac{pi}-based (PCRDI) and model-based (MCRDI) constraints, constrained \ac{rdi} can effectively eliminate oversubtraction for \ac{rdi} products. 
For the simple disk systems that can be feasibly modeled: PCRDI provides analysis-ready imagery orders of magnitude more quickly than the approximate corrections provided by forward-modeling, while MCRDI provides final products of higher quality and which are more conducive to detailed analysis of disk features and properties ({with both also allowing more accurate measurement of disk brightness and color than with model-based corrections}). For significantly extended and highly structured disks, PCRDI uniquely provides total intensity products which are unaffected by the significant and variable \ac{css} loss that normally inhibits studies of these systems. In turn, these products enable the detailed spectral and spatial analysis needed to robustly identify planets embedded in disks, like AB Aur b \citep{Currie2022}, and to conduct detailed studies of disk composition and morphology.   

Proliferation of these techniques, or others that yield comparable results, is paramount for realizing the capability of current high-contrast imaging systems. With the recent launch of \ac{jwst}, and with many high-impact direct imaging missions on the horizon -- including the Roman Space Telescope, and the observatories recommended by the 2020 Astronomy and Astrophysics Decadal Survey (the 6-meter LUVOIR/HabEx ``hybrid", and 30-meter ground-based telescopes), it is more important still to ensure that available post-processing tools are not the limiting factor in the yield of exoplanet and disk studies.

Software supporting the application of these techniques is planned for release in late 2022 (see Appendix \ref{app:runtimes}).

\newpage
\acknowledgements
{We thank our referee, whose comments helped us to improve both the content and clarity of this manuscript.}

This research is based on data collected at Subaru Telescope, which is operated by the National Astronomical Observatory of Japan. We are honored and grateful for the opportunity of observing the Universe from Maunakea, which has cultural, historical and natural significance in Hawaii.

We wish to acknowledge the critical importance of the current and recent Subaru telescope operators, daycrew, computer support, and office staff employees.  Their expertise, ingenuity, and dedication is indispensable to the continued successful operation of Subaru.  

The development of SCExAO was supported by the Japan Society for the Promotion of Science (Grant-in-Aid for Research \#23340051, \#26220704, \#23103002, \#19H00703 \& \#19H00695), the Astrobiology Center of the National Institutes of Natural Sciences, Japan, the Mt Cuba Foundation and the director’s contingency fund at Subaru Telescope.  We acknowledge funding support from the NASA XRP program via grants 80NSSC20K0252 and NNX17AF88G.  T.C. was supported by a NASA Senior Postdoctoral Fellowship. 

This work was directly enabled by \textit{Sigma Xi Grants In Aid of Research (GIAR)}, which provided support for the GPU-equipped workstation used in formulating the techniques presented here.

\software{Matplotlib \citep{matplotlib2007, matplotlib2021},
NumPy \citep{numpy2020},
SciPy \citep{scipy2020},
Astropy \citep{astropy2013, astropy2018},
CuPy \citep{cupy2017},
LMFIT \citep{Newville2014},
diskmap \citep{Stolker2016},
DiskDyn \citep{Gaspar2020},
WebbPSF \citep{Perrin2014},
CHARIS Data Reduction Pipeline \citep{Brandt2017},
CHARIS Data Processing Pipeline \citep{Currie2020spie}}

\newpage
\appendix
\section{Model-based RDI Corrections}\label{app:rdi_corrections}
{RDI forward modeling for a single target exposure is carried out as follows. Let $I_M$ be a synthetic disk model which has been rotated to the orientation of the target image and convolved with the \ac{psf}. Let the corresponding forward modeled result be $I_M^\prime$. For a conventional RDI procedure, $I_M^\prime$ can be found simply by substituting $I_M$ in place of $I_C$ in Eq. \ref{eq:rdi_atten}:}

\begin{equation}\label{eq:rdi_fwdmod}
I_M^\prime = I_M - \mathcal{M}(I_M, \mathcal{R})
\end{equation}

{Both $I_M^\prime$ and $I_M$ are then derotated to align with north-up. This process is repeated for each target image to create sequences of input (unattenuated) model images and processed (attenuated) model images matching the data sequence. }

{Typically, the input and processed model sequences are each averaged in the same manner as the real result to get final input and processed model images, whose ratio forms an attenuation correction map \citep[e.g.,][]{Goebel2018,Lawson2021b,Betti2022} \footnote{Some differences in the procedure are common to improve efficiency for data with a predominantly symmetric PSF.} by which the RDI result is multiplied to produce a model-corrected result. For our purposes, we instead compute attenuation corrections for each exposure in the target sequence --- dividing the full input model sequence by the output model sequence. In regions where the processed model image is very close to zero (e.g., when transitioning from positive to negative), large spurious values manifest in the corrections. To handle this, any correction factors with absolute values larger than 100 were replaced with a value of one. This does not impinge on the non-spurious corrections (which are predominantly $\lesssim 5$) and significantly improves the accuracy of SB measurements for the corrected RDI results. Limited testing with other values between 10 and 500 showed no significant impact on the accuracy of the corrected results.}

{Once prepared, we multiply the PSF subtracted and derotated data sequence with this correction sequence to form a sequence of model-corrected images. We then average this corrected sequence in the same manner as the uncorrected RDI result to reach the final model-corrected result. From testing with the simulated data of Section \ref{sec:pcrdi_sims}, we found that using exposure-by-exposure corrections --- rather than a single averaged correction --- improved the accuracy of the final results while also significantly reducing noise that otherwise remained in faint areas.}

\section{Surface Brightness Uncertainty}\label{app:sb_uncertainty}
{AB Aur and similar disks present a number of challenges when discussing signal to noise ratios and similar measurements of significance. Widely used methods for estimating the uncertainty of flux or surface brightness measurements use the variance among the set of values in an image having the same stellocentric separation \citep[e.g.,][]{Mawet2014, Lawson2021b}. For disks that fill much of the \ac{fov} (or all of it, as is the case for the CHARIS AB Aur data), there is real and significant astrophysical variation in the values at a given separation which will inflate such noise estimates. Since our reductions have different amounts of astrophysical throughput, such noise estimates would be inflated by differing amounts, and so it is extremely challenging to make a fair comparison of significance between different reductions with these techniques.}
    
{To estimate uncertainty of surface brightness (SB) and color profiles for our results, we adopt a strategy of statistical bootstrapping \citep{Efron1994} as follows. The final image for a reduction is normally created by taking the average of the sequence of $N_T$ target exposures (following PSF subtraction and derotation). Here, we instead form a ``final" product by averaging a random selection, with replacement, of $N_T$ images from the same sequence. We then carry out the relevant SB measurements on this result. This is repeated 10000 times in each case, retaining each set of measurements. The uncertainty for a measurement at a given position is then measured from the distribution among the 10000 bootstrapped results at that location. To ensure a fair comparison, each reduction being compared uses the same selection of frames for a given iteration. Since both PCRDI results (AB Aur and the simulated test case) use a number of exposures well in excess of the 50 samples recommended by \citet{Efron1994}), bootstrapping should provide at least comparable (between reductions) estimates of uncertainty for these measurements. However, as there is temporal correlation in any residual speckle noise, the variance between exposures for a specific location is not entirely uncorrelated --- meaning that the statistics derived this way are not fully robust. Nevertheless, testing with an RDI reduction of data without a disk (otherwise using the same reference data used in the simulations of Section \ref{sec:pcrdi_sims}) showed that noise estimated this way is, on average, within a factor of $\sim1.2$ of noise computed with the common radial method.}

\section{Forward Modeling for Constrained RDI}\label{app:pcrdi_fwdmod}
{For general use, formal forward modeling is not necessary for PCRDI products (and is mostly redundant for MCRDI); to assess geometry, the optimized PCRDI procedure typically provides sufficient throughput that products can simply be compared to \ac{psf}-convolved models. However, conventional forward modeling and forward-model-based flux corrections can still be used. Let $I_C^0$ be the PI-based estimate used with the real data, and let $I_M$ be a \ac{psf}-convolved synthetic disk model to be forward modeled. The corresponding forward modeled result can be computed as follows:} 

\begin{equation}\label{eq:pcrdi_fwdmod}
I_M^\prime = I_M - \mathcal{M}(I_M-I_C^0, \mathcal{R})
\end{equation}

{In other words, if $I_M$ is truly the underlying \ac{css} in the data, then $I_M - I_C^0$ is the amount of \ac{css} remaining in the data when we construct the stellar \ac{psf} model -- and thus determines the over (or under) subtraction that will occur.}

{Compared with conventional \ac{rdi} reductions using forward modeling strategies, PCRDI forward modeling maintains a number of distinct advantages. Since attenuation corrections will also scale any residual starlight or noise (see Section \ref{sec:pcrdi_sims}), mitigating the majority of oversubtraction with PCRDI will enable higher signal-to-noise in corrected products than in corrected RDI products. Additionally, being less affected by biasing from \ac{css} attenuation, the higher throughput PCRDI products can be used to make much better estimates of disk parameters, and so will likely require exploration of many fewer models overall.}

\begin{figure*}
    \centering
    \includegraphics[width=0.98\textwidth]{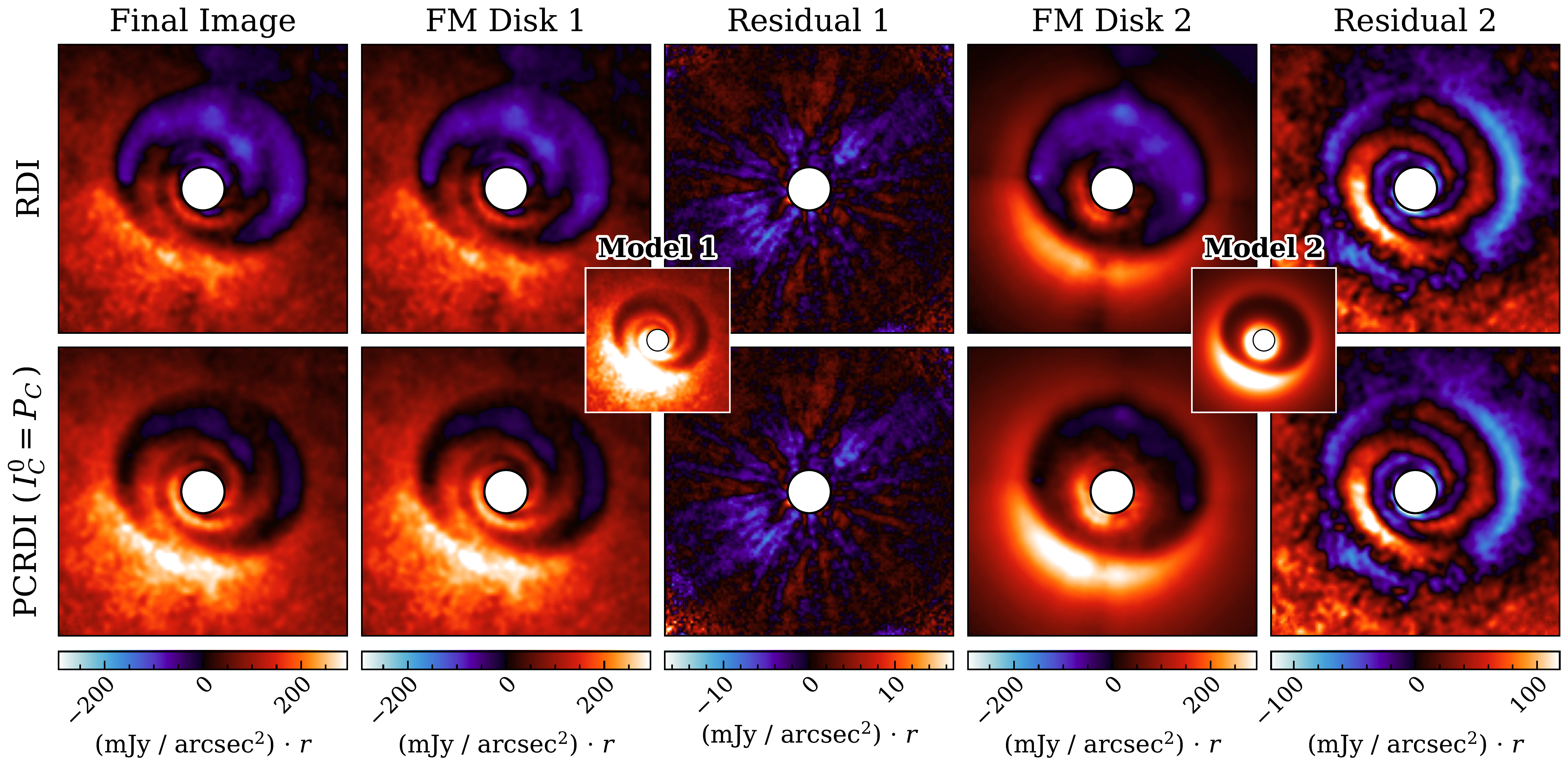}
    \caption{\textbf{Top row, left to right:} the standard RDI data result, model 1 (the ideal disk model) after RDI forward modeling, the residual map (data $-$ processed model) for model 1, model 2 (the non-ideal model) after forward modeling, and the residual map for model 2. \textbf{Bottom row:} As the top row, but for the non-optimized PCRDI reduction (directly adopting the PI image as the constraint instead of attempting to transform it to total intensity). The small inset images show the corresponding model before forward modeling. The residual columns, which are shown at a different color scale than the other columns for visibility, show very nearly identical residuals between the two reductions. 
    \label{fig:rdi_vs_pcrdi_fm}}
\end{figure*}

As an example, we compare forward modeled results for two reductions of the simulated data from Section \ref{sec:pcrdi_sims}: the standard \ac{rdi} result (as in Figure \ref{fig:pcrdi_sim_results}: top row, center-left panel), and a non-optimized PCRDI result (directly adopting the PI image as the \ac{css} estimate). For these reductions, we apply forward modeling to the two disk models previously used for RDI attenuation corrections in Section \ref{sec:pcrdi_sims}. These results are visualized in Figure \ref{fig:rdi_vs_pcrdi_fm} and demonstrate that --- even with the simplest constraint --- PCRDI significantly improves throughput while still permitting the same tools for interpretation of the results.

\section{Validation Using Conventional RDI Forward Modeling}\label{app:fwdmod}
In forward modeling, the effects of signal attenuation are induced on an input \ac{css} image. For disk studies, it is common to forward model \ac{psf}-convolved synthetic disk models to enable assessment of the geometry and properties of a real attenuated result; if a disk model can be identified which, when attenuation is induced, closely matches the real attenuated data, then that model provides a reasonable explanation of the true unattenuated disk signal. For a standard \ac{rdi} reduction, the attenuated result for a given input image can be found by evaluating Eq. \ref{eq:rdi_atten} with the input image in place of $I_C$. 

To further validate a final PCRDI result in a manner more familiar to the disk-imaging community, we can simply carry out this forward modeling procedure with the PCRDI result as the input image and then compare the result with the standard RDI product. If the attenuated RDI result is consistent with the PCRDI result when RDI attenuation is induced, then the PCRDI image provides a reasonable ``model'' for the unattenuated \ac{css}.

\begin{figure*}
    \centering
    \includegraphics[width=0.95\textwidth]{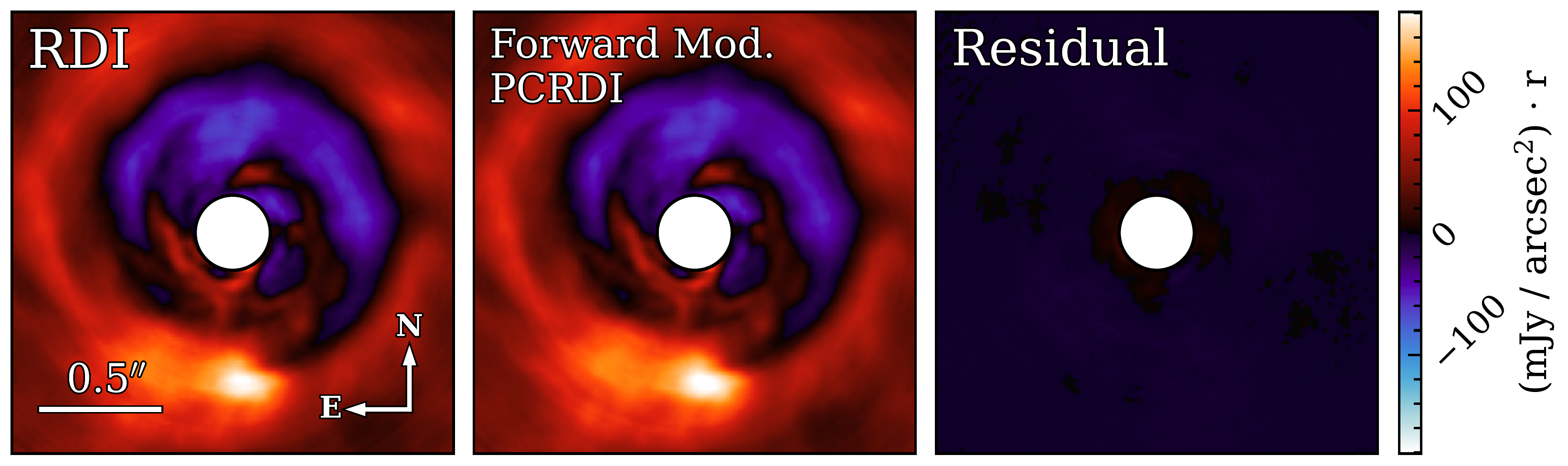}
    \caption{\textbf{Left:} the conventional RDI result for AB Aur (as in Figure \ref{fig:abaur_pcrdi}, top left). \textbf{Center:} the PCRDI result of Section \ref{sec:abaur} after inducing standard RDI attenuation via forward modeling. \textbf{Right:} The difference between the RDI result and the forward-modeled PCRDI result, demonstrating that the PCRDI product is an accurate representation of the unattenuated \ac{css}. All images have been wavelength-averaged and multiplied by the stellocentric separation in units of arcseconds (as in Figure \ref{fig:abaur_pcrdi}).
    \label{fig:pcrdi_fm}}
\end{figure*}

To demonstrate, we carried this out using the PCRDI result for AB Aurigae (Section \ref{sec:abaur}). Figure \ref{fig:pcrdi_fm} shows the result of this procedure. The negligible residual signal in this case provides additional evidence that our PCRDI optimization converged to a final result that is consistent with the \ac{css} contained in the data. 

It is important to note that forward modeling the unconstrained RDI result in the same manner will also result in similarly small residuals; a forward modeled image being consistent with the RDI result only indicates that the difference between the input image and the RDI result is well approximated by some combination of reference images (or reference image eigenvectors, in the case of KLIP). In other words, a good agreement between a given forward modeled result and the RDI result can only indicate that the input image is \emph{a} suitable solution (rather than \emph{the} suitable solution)\footnote{For any non-azimuthally symmetric \ac{psf} for data with non-negligible field rotation, the forward modeled constrained RDI result should yield smaller residuals; an oversubtracted result will have the imprint of the derotated and sequence-averaged PSF, which will not be properly aligned with the PSF when rotating the ``model" to the parallactic angles of the data for forward modeling. In testing, the improvement was small but non-negligible.}. In general, this test will not provide any better assessment of a constrained RDI result than the objective function calculation used in constrained RDI optimization (Section \ref{sec:optimization}), but may still be useful, e.g., for demonstrating that constrained RDI was implemented correctly.

\section{Supporting Software and Run-time}\label{app:runtimes}
The software written to carry out optimization of PCRDI and MCRDI makes extensive use of vectorization, multi-threading, and GPU acceleration to carry out these procedures relatively quickly.

Public release of Python software to support generalized application of the techniques described here is planned for late 2022. Until then, groups interested in using this technique (or who are interested in other uses of accelerated data reduction for high-contrast imaging) are encouraged to contact the authors.

\subsection{Run-time}
To carry out optimization of PCRDI and MCRDI with the aforementioned software, a personal computer with the following specifications was used:

\noindent{Memory: 31 GiB\\}
\noindent{Processor: Intel Core i9-10900X CPU @ 3.70GHz $\times$ 20\\}
\noindent{GPU: NVIDIA GeForce RTX 3070\\}

The dimensions and run-times for each dataset utilized are provided in Table \ref{tab:dataruntimes}.  Note: the number of function evaluations for optimization is generally detached from disk complexity for PCRDI, but is strongly linked for MCRDI (e.g., with a multi-ring system requiring optimization of more parameters).

\begin{deluxetable*}{@{\extracolsep{0pt}}cccccc}
    \tablewidth{0pt}
    \tablecaption{PCRDI and MCRDI Dataset Dimensionality and Optimization Run-times}
    \tablehead{
    \colhead{Dataset} & \colhead{$N_{\rm tar}$} & \colhead{$N_{\rm ref}$} & \colhead{Exposure dimensions} &  \colhead{Opt. $N_{\rm fev}$} & \colhead{Opt. time (s)}
    }
    \startdata
    PCRDI (sim. data) & 107 & 36 & [22 $\lambda$ $\times$ 201 pix $\times$ 201 pix] & 92 & 125 \\
    PCRDI (AB Aur) & 168 & 51 & [22 $\lambda$ $\times$ 201 pix $\times$ 201 pix] & 48 & 88 \\
    MCRDI (sim. data) & 30 & 15 & [320 pix $\times$ 320 pix] & 90 & 29 \\
    \enddata
    \tablecomments{For each dataset utilized in this work, the dimensions and total time required to reach the final constrained \ac{rdi} product. $N_{\rm tar}$ and $N_{\rm ref}$ give the number of target and reference exposures, respectively -- with the size of each image or image-cube given by ``exposure dimensions". ``Opt. $N_{\rm fev}$" and ``Opt. time" indicate the number of function evaluations and total run-time (in seconds) required to reach the presented solution, respectively.
    }\label{tab:dataruntimes}
\end{deluxetable*}

\bibliography{refs}{}

\begin{thebibliography}{}
\expandafter\ifx\csname natexlab\endcsname\relax\def\natexlab#1{#1}\fi
\providecommand{\url}[1]{\href{#1}{#1}}
\providecommand{\dodoi}[1]{doi:~\href{http://doi.org/#1}{\nolinkurl{#1}}}
\providecommand{\doeprint}[1]{\href{http://ascl.net/#1}{\nolinkurl{http://ascl.net/#1}}}
\providecommand{\doarXiv}[1]{\href{https://arxiv.org/abs/#1}{\nolinkurl{https://arxiv.org/abs/#1}}}

\bibitem[{{Astropy Collaboration} {et~al.}(2013){Astropy Collaboration},
  {Robitaille}, {Tollerud}, {Greenfield}, {Droettboom}, {Bray}, {Aldcroft},
  {Davis}, {Ginsburg}, {Price-Whelan}, {Kerzendorf}, {Conley}, {Crighton},
  {Barbary}, {Muna}, {Ferguson}, {Grollier}, {Parikh}, {Nair}, {Unther},
  {Deil}, {Woillez}, {Conseil}, {Kramer}, {Turner}, {Singer}, {Fox}, {Weaver},
  {Zabalza}, {Edwards}, {Azalee Bostroem}, {Burke}, {Casey}, {Crawford},
  {Dencheva}, {Ely}, {Jenness}, {Labrie}, {Lim}, {Pierfederici}, {Pontzen},
  {Ptak}, {Refsdal}, {Servillat}, \& {Streicher}}]{astropy2013}
{Astropy Collaboration}, {Robitaille}, T.~P., {Tollerud}, E.~J., {et~al.} 2013,
  \aap, 558, A33, \dodoi{10.1051/0004-6361/201322068}

\bibitem[{{Astropy Collaboration} {et~al.}(2018){Astropy Collaboration},
  {Price-Whelan}, {Sip{\H{o}}cz}, {G{\"u}nther}, {Lim}, {Crawford}, {Conseil},
  {Shupe}, {Craig}, {Dencheva}, {Ginsburg}, {VanderPlas}, {Bradley},
  {P{\'e}rez-Su{\'a}rez}, {de Val-Borro}, {Aldcroft}, {Cruz}, {Robitaille},
  {Tollerud}, {Ardelean}, {Babej}, {Bach}, {Bachetti}, {Bakanov}, {Bamford},
  {Barentsen}, {Barmby}, {Baumbach}, {Berry}, {Biscani}, {Boquien}, {Bostroem},
  {Bouma}, {Brammer}, {Bray}, {Breytenbach}, {Buddelmeijer}, {Burke},
  {Calderone}, {Cano Rodr{\'\i}guez}, {Cara}, {Cardoso}, {Cheedella}, {Copin},
  {Corrales}, {Crichton}, {D'Avella}, {Deil}, {Depagne}, {Dietrich}, {Donath},
  {Droettboom}, {Earl}, {Erben}, {Fabbro}, {Ferreira}, {Finethy}, {Fox},
  {Garrison}, {Gibbons}, {Goldstein}, {Gommers}, {Greco}, {Greenfield},
  {Groener}, {Grollier}, {Hagen}, {Hirst}, {Homeier}, {Horton}, {Hosseinzadeh},
  {Hu}, {Hunkeler}, {Ivezi{\'c}}, {Jain}, {Jenness}, {Kanarek}, {Kendrew},
  {Kern}, {Kerzendorf}, {Khvalko}, {King}, {Kirkby}, {Kulkarni}, {Kumar},
  {Lee}, {Lenz}, {Littlefair}, {Ma}, {Macleod}, {Mastropietro}, {McCully},
  {Montagnac}, {Morris}, {Mueller}, {Mumford}, {Muna}, {Murphy}, {Nelson},
  {Nguyen}, {Ninan}, {N{\"o}the}, {Ogaz}, {Oh}, {Parejko}, {Parley}, {Pascual},
  {Patil}, {Patil}, {Plunkett}, {Prochaska}, {Rastogi}, {Reddy Janga},
  {Sabater}, {Sakurikar}, {Seifert}, {Sherbert}, {Sherwood-Taylor}, {Shih},
  {Sick}, {Silbiger}, {Singanamalla}, {Singer}, {Sladen}, {Sooley},
  {Sornarajah}, {Streicher}, {Teuben}, {Thomas}, {Tremblay}, {Turner},
  {Terr{\'o}n}, {van Kerkwijk}, {de la Vega}, {Watkins}, {Weaver}, {Whitmore},
  {Woillez}, {Zabalza}, \& {Astropy Contributors}}]{astropy2018}
{Astropy Collaboration}, {Price-Whelan}, A.~M., {Sip{\H{o}}cz}, B.~M., {et~al.}
  2018, \aj, 156, 123, \dodoi{10.3847/1538-3881/aabc4f}

\bibitem[{{Augereau} {et~al.}(1999){Augereau}, {Lagrange}, {Mouillet},
  {Papaloizou}, \& {Grorod}}]{Augereau1999}
{Augereau}, J.~C., {Lagrange}, A.~M., {Mouillet}, D., {Papaloizou}, J.~C.~B.,
  \& {Grorod}, P.~A. 1999, \aap, 348, 557.
\newblock \doarXiv{astro-ph/9906429}

\bibitem[{{Avenhaus} {et~al.}(2018){Avenhaus}, {Quanz}, {Garufi}, {Perez},
  {Casassus}, {Pinte}, {Bertrang}, {Caceres}, {Benisty}, \&
  {Dominik}}]{Avenhaus2018}
{Avenhaus}, H., {Quanz}, S.~P., {Garufi}, A., {et~al.} 2018, \apj, 863, 44,
  \dodoi{10.3847/1538-4357/aab846}

\bibitem[{{Benisty} {et~al.}(2022){Benisty}, {Dominik}, {Follette}, {Garufi},
  {Ginski}, {Hashimoto}, {Keppler}, {Kley}, \& {Monnier}}]{Benisty2022}
{Benisty}, M., {Dominik}, C., {Follette}, K., {et~al.} 2022, arXiv e-prints,
  arXiv:2203.09991.
\newblock \doarXiv{2203.09991}

\bibitem[{{Betti} {et~al.}(2022){Betti}, {Follette}, {Jorquera}, {Duch{\^e}ne},
  {Mazoyer}, {Bonnefoy}, {Chauvin}, {P{\'e}rez}, {Boccaletti}, {Pinte},
  {Weinberger}, {Grady}, {Close}, {Defr{\`e}re}, {Downey}, {Hinz},
  {M{\'e}nard}, {Schneider}, {Skemer}, \& {Vaz}}]{Betti2022}
{Betti}, S.~K., {Follette}, K., {Jorquera}, S., {et~al.} 2022, arXiv e-prints,
  arXiv:2201.08868.
\newblock \doarXiv{2201.08868}

\bibitem[{{Beuzit} {et~al.}(2019){Beuzit}, {Vigan}, {Mouillet}, {Dohlen},
  {Gratton}, {Boccaletti}, {Sauvage}, {Schmid}, {Langlois}, {Petit},
  {Baruffolo}, {Feldt}, {Milli}, {Wahhaj}, {Abe}, {Anselmi}, {Antichi},
  {Barette}, {Baudrand}, {Baudoz}, {Bazzon}, {Bernardi}, {Blanchard}, {Brast},
  {Bruno}, {Buey}, {Carbillet}, {Carle}, {Cascone}, {Chapron}, {Charton},
  {Chauvin}, {Claudi}, {Costille}, {De Caprio}, {de Boer}, {Delboulb{\'e}},
  {Desidera}, {Dominik}, {Downing}, {Dupuis}, {Fabron}, {Fantinel}, {Farisato},
  {Feautrier}, {Fedrigo}, {Fusco}, {Gigan}, {Ginski}, {Girard}, {Giro},
  {Gisler}, {Gluck}, {Gry}, {Henning}, {Hubin}, {Hugot}, {Incorvaia}, {Jaquet},
  {Kasper}, {Lagadec}, {Lagrange}, {Le Coroller}, {Le Mignant}, {Le Ruyet},
  {Lessio}, {Lizon}, {Llored}, {Lundin}, {Madec}, {Magnard}, {Marteaud},
  {Martinez}, {Maurel}, {M{\'e}nard}, {Mesa}, {M{\"o}ller-Nilsson}, {Moulin},
  {Moutou}, {Orign{\'e}}, {Parisot}, {Pavlov}, {Perret}, {Pragt}, {Puget},
  {Rabou}, {Ramos}, {Reess}, {Rigal}, {Rochat}, {Roelfsema}, {Rousset}, {Roux},
  {Saisse}, {Salasnich}, {Santambrogio}, {Scuderi}, {Segransan}, {Sevin},
  {Siebenmorgen}, {Soenke}, {Stadler}, {Suarez}, {Tiph{\`e}ne}, {Turatto},
  {Udry}, {Vakili}, {Waters}, {Weber}, {Wildi}, {Zins}, \&
  {Zurlo}}]{Beuzit2019}
{Beuzit}, J.~L., {Vigan}, A., {Mouillet}, D., {et~al.} 2019, \aap, 631, A155,
  \dodoi{10.1051/0004-6361/201935251}

\bibitem[{{Bhowmik} {et~al.}(2019){Bhowmik}, {Boccaletti}, {Th{\'e}bault},
  {Kral}, {Mazoyer}, {Milli}, {Maire}, {van Holstein}, {Augereau}, {Baudoz},
  {Feldt}, {Galicher}, {Henning}, {Lagrange}, {Olofsson}, {Pantin}, \&
  {Perrot}}]{Bhowmik2019}
{Bhowmik}, T., {Boccaletti}, A., {Th{\'e}bault}, P., {et~al.} 2019, \aap, 630,
  A85, \dodoi{10.1051/0004-6361/201936076}

\bibitem[{{Boccaletti} {et~al.}(2020){Boccaletti}, {Di Folco}, {Pantin},
  {Dutrey}, {Guilloteau}, {Tang}, {Pi{\'e}tu}, {Habart}, {Milli}, {Beck}, \&
  {Maire}}]{Boccaletti2020}
{Boccaletti}, A., {Di Folco}, E., {Pantin}, E., {et~al.} 2020, \aap, 637, L5,
  \dodoi{10.1051/0004-6361/202038008}

\bibitem[{{Brandt} {et~al.}(2017){Brandt}, {Rizzo}, {Groff}, {Chilcote},
  {Greco}, {Kasdin}, {Limbach}, {Galvin}, {Loomis}, {Knapp}, {McElwain},
  {Jovanovic}, {Currie}, {Mede}, {Tamura}, {Takato}, \& {Hayashi}}]{Brandt2017}
{Brandt}, T.~D., {Rizzo}, M., {Groff}, T., {et~al.} 2017, Journal of
  Astronomical Telescopes, Instruments, and Systems, 3, 048002,
  \dodoi{10.1117/1.JATIS.3.4.048002}

\bibitem[{Caswell {et~al.}(2021)Caswell, Droettboom, Lee, de~Andrade, Hoffmann,
  Hunter, Klymak, Firing, Stansby, Varoquaux, Nielsen, Root, May, Elson,
  Seppänen, Dale, Lee, McDougall, Straw, Hobson, hannah, Gohlke, Vincent, Yu,
  Ma, Silvester, Moad, Kniazev, Ernest, \& Ivanov}]{matplotlib2021}
Caswell, T.~A., Droettboom, M., Lee, A., {et~al.} 2021, matplotlib/matplotlib:
  REL: v3.5.1, v3.5.1,  Zenodo, \dodoi{10.5281/zenodo.5773480}

\bibitem[{{Choquet} {et~al.}(2014){Choquet}, {Pueyo}, {Hagan}, {Gofas-Salas},
  {Rajan}, {Chen}, {Perrin}, {Debes}, {Golimowski}, {Hines}, {N'Diaye},
  {Schneider}, {Mawet}, {Marois}, \& {Soummer}}]{Choquet2014}
{Choquet}, {\'E}., {Pueyo}, L., {Hagan}, J.~B., {et~al.} 2014, in Society of
  Photo-Optical Instrumentation Engineers (SPIE) Conference Series, Vol. 9143,
  Space Telescopes and Instrumentation 2014: Optical, Infrared, and Millimeter
  Wave, ed. J.~{Oschmann}, Jacobus~M., M.~{Clampin}, G.~G. {Fazio}, \& H.~A.
  {MacEwen}, 914357, \dodoi{10.1117/12.2056672}

\bibitem[{{Currie} {et~al.}(2022{\natexlab{a}}){Currie}, {Biller}, {Lagrange},
  {Marois}, {Guyon}, {Nielsen}, {Bonnefoy}, \& {De Rosa}}]{Currie2022b}
{Currie}, T., {Biller}, B., {Lagrange}, A.-M., {et~al.} 2022{\natexlab{a}},
  arXiv e-prints, arXiv:2205.05696.
\newblock \doarXiv{2205.05696}

\bibitem[{{Currie} {et~al.}(2015){Currie}, {Cloutier}, {Brittain}, {Grady},
  {Burrows}, {Muto}, {Kenyon}, \& {Kuchner}}]{Currie2015}
{Currie}, T., {Cloutier}, R., {Brittain}, S., {et~al.} 2015, \apj, 814, L27,
  \dodoi{10.1088/2041-8205/814/2/L27}

\bibitem[{{Currie} {et~al.}(2012){Currie}, {Debes}, {Rodigas}, {Burrows},
  {Itoh}, {Fukagawa}, {Kenyon}, {Kuchner}, \& {Matsumura}}]{Currie2012}
{Currie}, T., {Debes}, J., {Rodigas}, T.~J., {et~al.} 2012, \apjl, 760, L32,
  \dodoi{10.1088/2041-8205/760/2/L32}

\bibitem[{{Currie} {et~al.}(2019){Currie}, {Marois}, {Cieza}, {Mulders},
  {Lawson}, {Caceres}, {Rodriguez-Ruiz}, {Wisniewski}, {Guyon}, {Brandt},
  {Kasdin}, {Groff}, {Lozi}, {Chilcote}, {Hodapp}, {Jovanovic}, {Martinache},
  {Skaf}, {Lyra}, {Tamura}, {Asensio-Torres}, {Dong}, {Grady}, {Gerard},
  {Fukagawa}, {Hand}, {Hayashi}, {Henning}, {Kudo}, {Kuzuhara}, {Kwon},
  {McElwain}, \& {Uyama}}]{Currie2019}
{Currie}, T., {Marois}, C., {Cieza}, L., {et~al.} 2019, \apjl, 877, L3,
  \dodoi{10.3847/2041-8213/ab1b42}

\bibitem[{{Currie} {et~al.}(2020){Currie}, {Guyon}, {Lozi}, {Sahoo}, {Vievard},
  {Deo}, {Chilcote}, {Groff}, {Brandt}, {Lawson}, {Skaf}, {Martinache}, \&
  {Kasdin}}]{Currie2020spie}
{Currie}, T., {Guyon}, O., {Lozi}, J., {et~al.} 2020, in Society of
  Photo-Optical Instrumentation Engineers (SPIE) Conference Series, Vol. 11448,
  Society of Photo-Optical Instrumentation Engineers (SPIE) Conference Series,
  114487H, \dodoi{10.1117/12.2576349}

\bibitem[{{Currie} {et~al.}(2022{\natexlab{b}}){Currie}, {Lawson}, {Schneider},
  {Lyra}, {Wisniewski}, {Grady}, {Guyon}, {Tamura}, {Kotani}, {Kawahara},
  {Brandt}, {Uyama}, {Muto}, {Dong}, {Kudo}, {Hashimoto}, {Fukagawa}, {Wagner},
  {Lozi}, {Chilcote}, {Tobin}, {Groff}, {Ward-Duong}, {Januszewski}, {Norris},
  {Tuthill}, {van der Marel}, {Sitko}, {Deo}, {Vievard}, {Jovanovic},
  {Martinache}, \& {Skaf}}]{Currie2022}
{Currie}, T., {Lawson}, K., {Schneider}, G., {et~al.} 2022{\natexlab{b}},
  Nature Astronomy, \dodoi{10.1038/s41550-022-01634-x}

\bibitem[{Efron {et~al.}(1994)Efron, Tibshirani, \& Tibshirani}]{Efron1994}
Efron, B., Tibshirani, R., \& Tibshirani, R.~J. 1994, An introduction to the
  bootstrap, Chapman \& Hall/CRC Monographs on Statistics and Applied
  Probability (Philadelphia, PA: Chapman \& Hall/CRC)

\bibitem[{{Esposito} {et~al.}(2020){Esposito}, {Kalas}, {Fitzgerald},
  {Millar-Blanchaer}, {Duch{\^e}ne}, {Patience}, {Hom}, {Perrin}, {De Rosa},
  {Chiang}, {Czekala}, {Macintosh}, {Graham}, {Ansdell}, {Arriaga}, {Bruzzone},
  {Bulger}, {Chen}, {Cotten}, {Dong}, {Draper}, {Follette}, {Hung}, {Lopez},
  {Matthews}, {Mazoyer}, {Metchev}, {Rameau}, {Ren}, {Rice}, {Song}, {Stahl},
  {Wang}, {Wolff}, {Zuckerman}, {Ammons}, {Bailey}, {Barman}, {Chilcote},
  {Doyon}, {Gerard}, {Goodsell}, {Greenbaum}, {Hibon}, {Hinkley}, {Ingraham},
  {Konopacky}, {Maire}, {Marchis}, {Marley}, {Marois}, {Nielsen},
  {Oppenheimer}, {Palmer}, {Poyneer}, {Pueyo}, {Rajan}, {Rantakyr{\"o}},
  {Ruffio}, {Savransky}, {Schneider}, {Sivaramakrishnan}, {Soummer}, {Thomas},
  \& {Ward-Duong}}]{Esposito2020}
{Esposito}, T.~M., {Kalas}, P., {Fitzgerald}, M.~P., {et~al.} 2020, \aj, 160,
  24, \dodoi{10.3847/1538-3881/ab9199}

\bibitem[{{Flasseur} {et~al.}(2018){Flasseur}, {Denis}, {Thi{\'e}baut}, \&
  {Langlois}}]{Flasseur2018}
{Flasseur}, O., {Denis}, L., {Thi{\'e}baut}, {\'E}., \& {Langlois}, M. 2018,
  \aap, 618, A138, \dodoi{10.1051/0004-6361/201832745}

\bibitem[{{Flasseur} {et~al.}(2021){Flasseur}, {Th{\'e}}, {Denis},
  {Thi{\'e}baut}, \& {Langlois}}]{Flasseur2021}
{Flasseur}, O., {Th{\'e}}, S., {Denis}, L., {Thi{\'e}baut}, {\'E}., \&
  {Langlois}, M. 2021, \aap, 651, A62, \dodoi{10.1051/0004-6361/202038957}

\bibitem[{{Fukagawa} {et~al.}(2004){Fukagawa}, {Hayashi}, {Tamura}, {Itoh},
  {Hayashi}, {Oasa}, {Takeuchi}, {Morino}, {Murakawa}, {Oya}, {Yamashita},
  {Suto}, {Mayama}, {Naoi}, {Ishii}, {Pyo}, {Nishikawa}, {Takato}, {Usuda},
  {Ando}, {Iye}, {Miyama}, \& {Kaifu}}]{Fukugawa2004}
{Fukagawa}, M., {Hayashi}, M., {Tamura}, M., {et~al.} 2004, \apjl, 605, L53,
  \dodoi{10.1086/420699}

\bibitem[{{Gaia Collaboration} {et~al.}(2016){Gaia Collaboration}, {Prusti},
  {de Bruijne}, {Brown}, {Vallenari}, {Babusiaux}, {Bailer-Jones}, {Bastian},
  {Biermann}, {Evans}, {Eyer}, {Jansen}, {Jordi}, {Klioner}, {Lammers},
  {Lindegren}, {Luri}, {Mignard}, {Milligan}, {Panem}, {Poinsignon},
  {Pourbaix}, {Randich}, {Sarri}, {Sartoretti}, {Siddiqui}, {Soubiran},
  {Valette}, {van Leeuwen}, {Walton}, {Aerts}, {Arenou}, {Cropper}, {Drimmel},
  {H{\o}g}, {Katz}, {Lattanzi}, {O'Mullane}, {Grebel}, {Holland}, {Huc},
  {Passot}, {Bramante}, {Cacciari}, {Casta{\~n}eda}, {Chaoul}, {Cheek}, {De
  Angeli}, {Fabricius}, {Guerra}, {Hern{\'a}ndez}, {Jean-Antoine-Piccolo},
  {Masana}, {Messineo}, {Mowlavi}, {Nienartowicz}, {Ord{\'o}{\~n}ez-Blanco},
  {Panuzzo}, {Portell}, {Richards}, {Riello}, {Seabroke}, {Tanga},
  {Th{\'e}venin}, {Torra}, {Els}, {Gracia-Abril}, {Comoretto},
  {Garcia-Reinaldos}, {Lock}, {Mercier}, {Altmann}, {Andrae}, {Astraatmadja},
  {Bellas-Velidis}, {Benson}, {Berthier}, {Blomme}, {Busso}, {Carry},
  {Cellino}, {Clementini}, {Cowell}, {Creevey}, {Cuypers}, {Davidson}, {De
  Ridder}, {de Torres}, {Delchambre}, {Dell'Oro}, {Ducourant}, {Fr{\'e}mat},
  {Garc{\'\i}a-Torres}, {Gosset}, {Halbwachs}, {Hambly}, {Harrison}, {Hauser},
  {Hestroffer}, {Hodgkin}, {Huckle}, {Hutton}, {Jasniewicz}, {Jordan},
  {Kontizas}, {Korn}, {Lanzafame}, {Manteiga}, {Moitinho}, {Muinonen},
  {Osinde}, {Pancino}, {Pauwels}, {Petit}, {Recio-Blanco}, {Robin}, {Sarro},
  {Siopis}, {Smith}, {Smith}, {Sozzetti}, {Thuillot}, {van Reeven}, {Viala},
  {Abbas}, {Abreu Aramburu}, {Accart}, {Aguado}, {Allan}, {Allasia},
  {Altavilla}, {{\'A}lvarez}, {Alves}, {Anderson}, {Andrei}, {Anglada Varela},
  {Antiche}, {Antoja}, {Ant{\'o}n}, {Arcay}, {Atzei}, {Ayache}, {Bach},
  {Baker}, {Balaguer-N{\'u}{\~n}ez}, {Barache}, {Barata}, {Barbier}, {Barblan},
  {Baroni}, {Barrado y Navascu{\'e}s}, {Barros}, {Barstow}, {Becciani},
  {Bellazzini}, {Bellei}, {Bello Garc{\'\i}a}, {Belokurov}, {Bendjoya},
  {Berihuete}, {Bianchi}, {Bienaym{\'e}}, {Billebaud}, {Blagorodnova},
  {Blanco-Cuaresma}, {Boch}, {Bombrun}, {Borrachero}, {Bouquillon}, {Bourda},
  {Bouy}, {Bragaglia}, {Breddels}, {Brouillet}, {Br{\"u}semeister},
  {Bucciarelli}, {Budnik}, {Burgess}, {Burgon}, {Burlacu}, {Busonero}, {Buzzi},
  {Caffau}, {Cambras}, {Campbell}, {Cancelliere}, {Cantat-Gaudin}, {Carlucci},
  {Carrasco}, {Castellani}, {Charlot}, {Charnas}, {Charvet}, {Chassat},
  {Chiavassa}, {Clotet}, {Cocozza}, {Collins}, {Collins}, {Costigan}, {Crifo},
  {Cross}, {Crosta}, {Crowley}, {Dafonte}, {Damerdji}, {Dapergolas}, {David},
  {David}, {De Cat}, {de Felice}, {de Laverny}, {De Luise}, {De March}, {de
  Martino}, {de Souza}, {Debosscher}, {del Pozo}, {Delbo}, {Delgado},
  {Delgado}, {di Marco}, {Di Matteo}, {Diakite}, {Distefano}, {Dolding}, {Dos
  Anjos}, {Drazinos}, {Dur{\'a}n}, {Dzigan}, {Ecale}, {Edvardsson}, {Enke},
  {Erdmann}, {Escolar}, {Espina}, {Evans}, {Eynard Bontemps}, {Fabre},
  {Fabrizio}, {Faigler}, {Falc{\~a}o}, {Farr{\`a}s Casas}, {Faye}, {Federici},
  {Fedorets}, {Fern{\'a}ndez-Hern{\'a}ndez}, {Fernique}, {Fienga}, {Figueras},
  {Filippi}, {Findeisen}, {Fonti}, {Fouesneau}, {Fraile}, {Fraser}, {Fuchs},
  {Furnell}, {Gai}, {Galleti}, {Galluccio}, {Garabato}, {Garc{\'\i}a-Sedano},
  {Gar{\'e}}, {Garofalo}, {Garralda}, {Gavras}, {Gerssen}, {Geyer}, {Gilmore},
  {Girona}, {Giuffrida}, {Gomes}, {Gonz{\'a}lez-Marcos},
  {Gonz{\'a}lez-N{\'u}{\~n}ez}, {Gonz{\'a}lez-Vidal}, {Granvik}, {Guerrier},
  {Guillout}, {Guiraud}, {G{\'u}rpide}, {Guti{\'e}rrez-S{\'a}nchez}, {Guy},
  {Haigron}, {Hatzidimitriou}, {Haywood}, {Heiter}, {Helmi}, {Hobbs},
  {Hofmann}, {Holl}, {Holland}, {Hunt}, {Hypki}, {Icardi}, {Irwin}, {Jevardat
  de Fombelle}, {Jofr{\'e}}, {Jonker}, {Jorissen}, {Julbe}, {Karampelas},
  {Kochoska}, {Kohley}, {Kolenberg}, {Kontizas}, {Koposov}, {Kordopatis},
  {Koubsky}, {Kowalczyk}, {Krone-Martins}, {Kudryashova}, {Kull}, {Bachchan},
  {Lacoste-Seris}, {Lanza}, {Lavigne}, {Le Poncin-Lafitte}, {Lebreton},
  {Lebzelter}, {Leccia}, {Leclerc}, {Lecoeur-Taibi}, {Lemaitre}, {Lenhardt},
  {Leroux}, {Liao}, {Licata}, {Lindstr{\o}m}, {Lister}, {Livanou}, {Lobel},
  {L{\"o}ffler}, {L{\'o}pez}, {Lopez-Lozano}, {Lorenz}, {Loureiro},
  {MacDonald}, {Magalh{\~a}es Fernandes}, {Managau}, {Mann}, {Mantelet},
  {Marchal}, {Marchant}, {Marconi}, {Marie}, {Marinoni}, {Marrese},
  {Marschalk{\'o}}, {Marshall}, {Mart{\'\i}n-Fleitas}, {Martino}, {Mary},
  {Matijevi{\v{c}}}, {Mazeh}, {McMillan}, {Messina}, {Mestre}, {Michalik},
  {Millar}, {Miranda}, {Molina}, {Molinaro}, {Molinaro}, {Moln{\'a}r},
  {Moniez}, {Montegriffo}, {Monteiro}, {Mor}, {Mora}, {Morbidelli}, {Morel},
  {Morgenthaler}, {Morley}, {Morris}, {Mulone}, {Muraveva}, {Musella},
  {Narbonne}, {Nelemans}, {Nicastro}, {Noval}, {Ord{\'e}novic},
  {Ordieres-Mer{\'e}}, {Osborne}, {Pagani}, {Pagano}, {Pailler}, {Palacin},
  {Palaversa}, {Parsons}, {Paulsen}, {Pecoraro}, {Pedrosa}, {Pentik{\"a}inen},
  {Pereira}, {Pichon}, {Piersimoni}, {Pineau}, {Plachy}, {Plum}, {Poujoulet},
  {Pr{\v{s}}a}, {Pulone}, {Ragaini}, {Rago}, {Rambaux}, {Ramos-Lerate},
  {Ranalli}, {Rauw}, {Read}, {Regibo}, {Renk}, {Reyl{\'e}}, {Ribeiro},
  {Rimoldini}, {Ripepi}, {Riva}, {Rixon}, {Roelens}, {Romero-G{\'o}mez},
  {Rowell}, {Royer}, {Rudolph}, {Ruiz-Dern}, {Sadowski}, {Sagrist{\`a}
  Sell{\'e}s}, {Sahlmann}, {Salgado}, {Salguero}, {Sarasso}, {Savietto},
  {Schnorhk}, {Schultheis}, {Sciacca}, {Segol}, {Segovia}, {Segransan},
  {Serpell}, {Shih}, {Smareglia}, {Smart}, {Smith}, {Solano}, {Solitro},
  {Sordo}, {Soria Nieto}, {Souchay}, {Spagna}, {Spoto}, {Stampa}, {Steele},
  {Steidelm{\"u}ller}, {Stephenson}, {Stoev}, {Suess}, {S{\"u}veges}, {Surdej},
  {Szabados}, {Szegedi-Elek}, {Tapiador}, {Taris}, {Tauran}, {Taylor},
  {Teixeira}, {Terrett}, {Tingley}, {Trager}, {Turon}, {Ulla}, {Utrilla},
  {Valentini}, {van Elteren}, {Van Hemelryck}, {van Leeuwen}, {Varadi},
  {Vecchiato}, {Veljanoski}, {Via}, {Vicente}, {Vogt}, {Voss}, {Votruba},
  {Voutsinas}, {Walmsley}, {Weiler}, {Weingrill}, {Werner}, {Wevers},
  {Whitehead}, {Wyrzykowski}, {Yoldas}, {{\v{Z}}erjal}, {Zucker}, {Zurbach},
  {Zwitter}, {Alecu}, {Allen}, {Allende Prieto}, {Amorim},
  {Anglada-Escud{\'e}}, {Arsenijevic}, {Azaz}, {Balm}, {Beck}, {Bernstein},
  {Bigot}, {Bijaoui}, {Blasco}, {Bonfigli}, {Bono}, {Boudreault}, {Bressan},
  {Brown}, {Brunet}, {Bunclark}, {Buonanno}, {Butkevich}, {Carret}, {Carrion},
  {Chemin}, {Ch{\'e}reau}, {Corcione}, {Darmigny}, {de Boer}, {de Teodoro}, {de
  Zeeuw}, {Delle Luche}, {Domingues}, {Dubath}, {Fodor}, {Fr{\'e}zouls},
  {Fries}, {Fustes}, {Fyfe}, {Gallardo}, {Gallegos}, {Gardiol}, {Gebran},
  {Gomboc}, {G{\'o}mez}, {Grux}, {Gueguen}, {Heyrovsky}, {Hoar}, {Iannicola},
  {Isasi Parache}, {Janotto}, {Joliet}, {Jonckheere}, {Keil}, {Kim},
  {Klagyivik}, {Klar}, {Knude}, {Kochukhov}, {Kolka}, {Kos}, {Kutka}, {Lainey},
  {LeBouquin}, {Liu}, {Loreggia}, {Makarov}, {Marseille}, {Martayan},
  {Martinez-Rubi}, {Massart}, {Meynadier}, {Mignot}, {Munari}, {Nguyen},
  {Nordlander}, {Ocvirk}, {O'Flaherty}, {Olias Sanz}, {Ortiz}, {Osorio},
  {Oszkiewicz}, {Ouzounis}, {Palmer}, {Park}, {Pasquato}, {Peltzer}, {Peralta},
  {P{\'e}turaud}, {Pieniluoma}, {Pigozzi}, {Poels}, {Prat}, {Prod'homme},
  {Raison}, {Rebordao}, {Risquez}, {Rocca-Volmerange}, {Rosen}, {Ruiz-Fuertes},
  {Russo}, {Sembay}, {Serraller Vizcaino}, {Short}, {Siebert}, {Silva},
  {Sinachopoulos}, {Slezak}, {Soffel}, {Sosnowska}, {Strai{\v{z}}ys}, {ter
  Linden}, {Terrell}, {Theil}, {Tiede}, {Troisi}, {Tsalmantza}, {Tur},
  {Vaccari}, {Vachier}, {Valles}, {Van Hamme}, {Veltz}, {Virtanen}, {Wallut},
  {Wichmann}, {Wilkinson}, {Ziaeepour}, \& {Zschocke}}]{Gaia2016}
{Gaia Collaboration}, {Prusti}, T., {de Bruijne}, J.~H.~J., {et~al.} 2016,
  \aap, 595, A1, \dodoi{10.1051/0004-6361/201629272}

\bibitem[{{Gaia Collaboration} {et~al.}(2021){Gaia Collaboration}, {Brown},
  {Vallenari}, {Prusti}, {de Bruijne}, {Babusiaux}, {Biermann}, {Creevey},
  {Evans}, {Eyer}, {Hutton}, {Jansen}, {Jordi}, {Klioner}, {Lammers},
  {Lindegren}, {Luri}, {Mignard}, {Panem}, {Pourbaix}, {Randich}, {Sartoretti},
  {Soubiran}, {Walton}, {Arenou}, {Bailer-Jones}, {Bastian}, {Cropper},
  {Drimmel}, {Katz}, {Lattanzi}, {van Leeuwen}, {Bakker}, {Cacciari},
  {Casta{\~n}eda}, {De Angeli}, {Ducourant}, {Fabricius}, {Fouesneau},
  {Fr{\'e}mat}, {Guerra}, {Guerrier}, {Guiraud}, {Jean-Antoine Piccolo},
  {Masana}, {Messineo}, {Mowlavi}, {Nicolas}, {Nienartowicz}, {Pailler},
  {Panuzzo}, {Riclet}, {Roux}, {Seabroke}, {Sordo}, {Tanga}, {Th{\'e}venin},
  {Gracia-Abril}, {Portell}, {Teyssier}, {Altmann}, {Andrae}, {Bellas-Velidis},
  {Benson}, {Berthier}, {Blomme}, {Brugaletta}, {Burgess}, {Busso}, {Carry},
  {Cellino}, {Cheek}, {Clementini}, {Damerdji}, {Davidson}, {Delchambre},
  {Dell'Oro}, {Fern{\'a}ndez-Hern{\'a}ndez}, {Galluccio}, {Garc{\'\i}a-Lario},
  {Garcia-Reinaldos}, {Gonz{\'a}lez-N{\'u}{\~n}ez}, {Gosset}, {Haigron},
  {Halbwachs}, {Hambly}, {Harrison}, {Hatzidimitriou}, {Heiter},
  {Hern{\'a}ndez}, {Hestroffer}, {Hodgkin}, {Holl}, {Jan{\ss}en}, {Jevardat de
  Fombelle}, {Jordan}, {Krone-Martins}, {Lanzafame}, {L{\"o}ffler}, {Lorca},
  {Manteiga}, {Marchal}, {Marrese}, {Moitinho}, {Mora}, {Muinonen}, {Osborne},
  {Pancino}, {Pauwels}, {Petit}, {Recio-Blanco}, {Richards}, {Riello},
  {Rimoldini}, {Robin}, {Roegiers}, {Rybizki}, {Sarro}, {Siopis}, {Smith},
  {Sozzetti}, {Ulla}, {Utrilla}, {van Leeuwen}, {van Reeven}, {Abbas}, {Abreu
  Aramburu}, {Accart}, {Aerts}, {Aguado}, {Ajaj}, {Altavilla}, {{\'A}lvarez},
  {{\'A}lvarez Cid-Fuentes}, {Alves}, {Anderson}, {Anglada Varela}, {Antoja},
  {Audard}, {Baines}, {Baker}, {Balaguer-N{\'u}{\~n}ez}, {Balbinot}, {Balog},
  {Barache}, {Barbato}, {Barros}, {Barstow}, {Bartolom{\'e}}, {Bassilana},
  {Bauchet}, {Baudesson-Stella}, {Becciani}, {Bellazzini}, {Bernet}, {Bertone},
  {Bianchi}, {Blanco-Cuaresma}, {Boch}, {Bombrun}, {Bossini}, {Bouquillon},
  {Bragaglia}, {Bramante}, {Breedt}, {Bressan}, {Brouillet}, {Bucciarelli},
  {Burlacu}, {Busonero}, {Butkevich}, {Buzzi}, {Caffau}, {Cancelliere},
  {C{\'a}novas}, {Cantat-Gaudin}, {Carballo}, {Carlucci}, {Carnerero},
  {Carrasco}, {Casamiquela}, {Castellani}, {Castro-Ginard}, {Castro Sampol},
  {Chaoul}, {Charlot}, {Chemin}, {Chiavassa}, {Cioni}, {Comoretto}, {Cooper},
  {Cornez}, {Cowell}, {Crifo}, {Crosta}, {Crowley}, {Dafonte}, {Dapergolas},
  {David}, {David}, {de Laverny}, {De Luise}, {De March}, {De Ridder}, {de
  Souza}, {de Teodoro}, {de Torres}, {del Peloso}, {del Pozo}, {Delbo},
  {Delgado}, {Delgado}, {Delisle}, {Di Matteo}, {Diakite}, {Diener},
  {Distefano}, {Dolding}, {Eappachen}, {Edvardsson}, {Enke}, {Esquej}, {Fabre},
  {Fabrizio}, {Faigler}, {Fedorets}, {Fernique}, {Fienga}, {Figueras},
  {Fouron}, {Fragkoudi}, {Fraile}, {Franke}, {Gai}, {Garabato},
  {Garcia-Gutierrez}, {Garc{\'\i}a-Torres}, {Garofalo}, {Gavras}, {Gerlach},
  {Geyer}, {Giacobbe}, {Gilmore}, {Girona}, {Giuffrida}, {Gomel}, {Gomez},
  {Gonzalez-Santamaria}, {Gonz{\'a}lez-Vidal}, {Granvik},
  {Guti{\'e}rrez-S{\'a}nchez}, {Guy}, {Hauser}, {Haywood}, {Helmi}, {Hidalgo},
  {Hilger}, {H{\l}adczuk}, {Hobbs}, {Holland}, {Huckle}, {Jasniewicz},
  {Jonker}, {Juaristi Campillo}, {Julbe}, {Karbevska}, {Kervella}, {Khanna},
  {Kochoska}, {Kontizas}, {Kordopatis}, {Korn}, {Kostrzewa-Rutkowska},
  {Kruszy{\'n}ska}, {Lambert}, {Lanza}, {Lasne}, {Le Campion}, {Le Fustec},
  {Lebreton}, {Lebzelter}, {Leccia}, {Leclerc}, {Lecoeur-Taibi}, {Liao},
  {Licata}, {Lindstr{\o}m}, {Lister}, {Livanou}, {Lobel}, {Madrero Pardo},
  {Managau}, {Mann}, {Marchant}, {Marconi}, {Marcos Santos}, {Marinoni},
  {Marocco}, {Marshall}, {Martin Polo}, {Mart{\'\i}n-Fleitas}, {Masip},
  {Massari}, {Mastrobuono-Battisti}, {Mazeh}, {McMillan}, {Messina},
  {Michalik}, {Millar}, {Mints}, {Molina}, {Molinaro}, {Moln{\'a}r},
  {Montegriffo}, {Mor}, {Morbidelli}, {Morel}, {Morris}, {Mulone}, {Munoz},
  {Muraveva}, {Murphy}, {Musella}, {Noval}, {Ord{\'e}novic}, {Orr{\`u}},
  {Osinde}, {Pagani}, {Pagano}, {Palaversa}, {Palicio}, {Panahi}, {Pawlak},
  {Pe{\~n}alosa Esteller}, {Penttil{\"a}}, {Piersimoni}, {Pineau}, {Plachy},
  {Plum}, {Poggio}, {Poretti}, {Poujoulet}, {Pr{\v{s}}a}, {Pulone}, {Racero},
  {Ragaini}, {Rainer}, {Raiteri}, {Rambaux}, {Ramos}, {Ramos-Lerate}, {Re
  Fiorentin}, {Regibo}, {Reyl{\'e}}, {Ripepi}, {Riva}, {Rixon}, {Robichon},
  {Robin}, {Roelens}, {Rohrbasser}, {Romero-G{\'o}mez}, {Rowell}, {Royer},
  {Rybicki}, {Sadowski}, {Sagrist{\`a} Sell{\'e}s}, {Sahlmann}, {Salgado},
  {Salguero}, {Samaras}, {Sanchez Gimenez}, {Sanna}, {Santove{\~n}a},
  {Sarasso}, {Schultheis}, {Sciacca}, {Segol}, {Segovia}, {S{\'e}gransan},
  {Semeux}, {Shahaf}, {Siddiqui}, {Siebert}, {Siltala}, {Slezak}, {Smart},
  {Solano}, {Solitro}, {Souami}, {Souchay}, {Spagna}, {Spoto}, {Steele},
  {Steidelm{\"u}ller}, {Stephenson}, {S{\"u}veges}, {Szabados}, {Szegedi-Elek},
  {Taris}, {Tauran}, {Taylor}, {Teixeira}, {Thuillot}, {Tonello}, {Torra},
  {Torra}, {Turon}, {Unger}, {Vaillant}, {van Dillen}, {Vanel}, {Vecchiato},
  {Viala}, {Vicente}, {Voutsinas}, {Weiler}, {Wevers}, {Wyrzykowski}, {Yoldas},
  {Yvard}, {Zhao}, {Zorec}, {Zucker}, {Zurbach}, \& {Zwitter}}]{Gaia2021}
{Gaia Collaboration}, {Brown}, A.~G.~A., {Vallenari}, A., {et~al.} 2021, \aap,
  649, A1, \dodoi{10.1051/0004-6361/202039657}

\bibitem[{{Garufi} {et~al.}(2022){Garufi}, {Dominik}, {Ginski}, {Benisty}, {van
  Holstein}, {Henning}, {Pawellek}, {Pinte}, {Avenhaus}, {Facchini},
  {Galicher}, {Gratton}, {M{\'e}nard}, {Muro-Arena}, {Milli}, {Stolker},
  {Vigan}, {Villenave}, {Moulin}, {Origne}, {Rigal}, {Sauvage}, \&
  {Weber}}]{Garufi2022}
{Garufi}, A., {Dominik}, C., {Ginski}, C., {et~al.} 2022, \aap, 658, A137,
  \dodoi{10.1051/0004-6361/202141692}

\bibitem[{{Gaspar} \& {Rieke}(2020)}]{Gaspar2020}
{Gaspar}, A., \& {Rieke}, G. 2020, Proceedings of the National Academy of
  Science, 117, 9712

\bibitem[{{Goebel} {et~al.}(2018){Goebel}, {Currie}, {Guyon}, {Brand t},
  {Groff}, {Jovanovic}, {Kasdin}, {Lozi}, {Hodapp}, {Martinache}, {Grady},
  {Hayashi}, {Kwon}, {McElwain}, {Yang}, \& {Tamura}}]{Goebel2018}
{Goebel}, S., {Currie}, T., {Guyon}, O., {et~al.} 2018, \aj, 156, 279,
  \dodoi{10.3847/1538-3881/aaeb24}

\bibitem[{{Gomez Gonzalez} {et~al.}(2017){Gomez Gonzalez}, {Wertz}, {Absil},
  {Christiaens}, {Defr{\`e}re}, {Mawet}, {Milli}, {Absil}, {Van Droogenbroeck},
  {Cantalloube}, {Hinz}, {Skemer}, {Karlsson}, \& {Surdej}}]{Gonzalez2017}
{Gomez Gonzalez}, C.~A., {Wertz}, O., {Absil}, O., {et~al.} 2017, \aj, 154, 7,
  \dodoi{10.3847/1538-3881/aa73d7}

\bibitem[{{Grady} {et~al.}(1999){Grady}, {Woodgate}, {Bruhweiler}, {Boggess},
  {Plait}, {Lindler}, {Clampin}, \& {Kalas}}]{Grady1999}
{Grady}, C.~A., {Woodgate}, B., {Bruhweiler}, F.~C., {et~al.} 1999, \apjl, 523,
  L151, \dodoi{10.1086/312270}

\bibitem[{{Groff} {et~al.}(2016){Groff}, {Chilcote}, {Kasdin}, {Galvin},
  {Loomis}, {Carr}, {Brand t}, {Knapp}, {Limbach}, {Guyon}, {Jovanovic},
  {McElwain}, {Takato}, \& {Hayashi}}]{Groff2016}
{Groff}, T.~D., {Chilcote}, J., {Kasdin}, N.~J., {et~al.} 2016, in Society of
  Photo-Optical Instrumentation Engineers (SPIE) Conference Series, Vol. 9908,
  Ground-based and Airborne Instrumentation for Astronomy VI, 99080O,
  \dodoi{10.1117/12.2233447}

\bibitem[{{Hagan} {et~al.}(2018){Hagan}, {Choquet}, {Soummer}, \&
  {Vigan}}]{Hagan2018}
{Hagan}, J.~B., {Choquet}, {\'E}., {Soummer}, R., \& {Vigan}, A. 2018, \aj,
  155, 179, \dodoi{10.3847/1538-3881/aab14b}

\bibitem[{Harris {et~al.}(2020)Harris, Millman, van~der Walt, Gommers,
  Virtanen, Cournapeau, Wieser, Taylor, Berg, Smith, Kern, Picus, Hoyer, van
  Kerkwijk, Brett, Haldane, del R{\'{i}}o, Wiebe, Peterson,
  G{\'{e}}rard-Marchant, Sheppard, Reddy, Weckesser, Abbasi, Gohlke, \&
  Oliphant}]{numpy2020}
Harris, C.~R., Millman, K.~J., van~der Walt, S.~J., {et~al.} 2020, Nature, 585,
  357, \dodoi{10.1038/s41586-020-2649-2}

\bibitem[{{Hashimoto} {et~al.}(2011){Hashimoto}, {Tamura}, {Muto}, {Kudo},
  {Fukagawa}, {Fukue}, {Goto}, {Grady}, {Henning}, {Hodapp}, {Honda},
  {Inutsuka}, {Kokubo}, {Knapp}, {McElwain}, {Momose}, {Ohashi}, {Okamoto},
  {Takami}, {Turner}, {Wisniewski}, {Janson}, {Abe}, {Brandner}, {Carson},
  {Egner}, {Feldt}, {Golota}, {Guyon}, {Hayano}, {Hayashi}, {Hayashi}, {Ishii},
  {Kandori}, {Kusakabe}, {Matsuo}, {Mayama}, {Miyama}, {Morino}, {Moro-Martin},
  {Nishimura}, {Pyo}, {Suto}, {Suzuki}, {Takato}, {Terada}, {Thalmann},
  {Tomono}, {Watanabe}, {Yamada}, {Takami}, \& {Usuda}}]{Hashimoto2011}
{Hashimoto}, J., {Tamura}, M., {Muto}, T., {et~al.} 2011, \apjl, 729, L17,
  \dodoi{10.1088/2041-8205/729/2/L17}

\bibitem[{{Hong}(1985)}]{Hong1985}
{Hong}, S.~S. 1985, \aap, 146, 67

\bibitem[{Hunter(2007)}]{matplotlib2007}
Hunter, J.~D. 2007, Computing in Science \& Engineering, 9, 90,
  \dodoi{10.1109/MCSE.2007.55}

\bibitem[{{Jovanovic} {et~al.}(2015){Jovanovic}, {Martinache}, {Guyon},
  {Clergeon}, {Singh}, {Kudo}, {Garrel}, {Newman}, {Doughty}, {Lozi}, {Males},
  {Minowa}, {Hayano}, {Takato}, {Morino}, {Kuhn}, {Serabyn}, {Norris},
  {Tuthill}, {Schworer}, {Stewart}, {Close}, {Huby}, {Perrin}, {Lacour},
  {Gauchet}, {Vievard}, {Murakami}, {Oshiyama}, {Baba}, {Matsuo}, {Nishikawa},
  {Tamura}, {Lai}, {Marchis}, {Duchene}, {Kotani}, \&
  {Woillez}}]{Jovanovic2015}
{Jovanovic}, N., {Martinache}, F., {Guyon}, O., {et~al.} 2015, \pasp, 127, 890,
  \dodoi{10.1086/682989}

\bibitem[{{Kenyon} {et~al.}(2008){Kenyon}, {G{\'o}mez}, \&
  {Whitney}}]{Kenyon2008taurusauriga}
{Kenyon}, S.~J., {G{\'o}mez}, M., \& {Whitney}, B.~A. 2008, in Handbook of Star
  Forming Regions, Volume I, ed. B.~{Reipurth}, Vol.~4, 405

\bibitem[{{Keppler} {et~al.}(2018){Keppler}, {Benisty}, {M{\"u}ller},
  {Henning}, {van Boekel}, {Cantalloube}, {Ginski}, {van Holstein}, {Maire},
  {Pohl}, {Samland}, {Avenhaus}, {Baudino}, {Boccaletti}, {de Boer},
  {Bonnefoy}, {Chauvin}, {Desidera}, {Langlois}, {Lazzoni}, {Marleau},
  {Mordasini}, {Pawellek}, {Stolker}, {Vigan}, {Zurlo}, {Birnstiel},
  {Brandner}, {Feldt}, {Flock}, {Girard}, {Gratton}, {Hagelberg}, {Isella},
  {Janson}, {Juhasz}, {Kemmer}, {Kral}, {Lagrange}, {Launhardt}, {Matter},
  {M{\'e}nard}, {Milli}, {Molli{\`e}re}, {Olofsson}, {P{\'e}rez}, {Pinilla},
  {Pinte}, {Quanz}, {Schmidt}, {Udry}, {Wahhaj}, {Williams}, {Buenzli},
  {Cudel}, {Dominik}, {Galicher}, {Kasper}, {Lannier}, {Mesa}, {Mouillet},
  {Peretti}, {Perrot}, {Salter}, {Sissa}, {Wildi}, {Abe}, {Antichi},
  {Augereau}, {Baruffolo}, {Baudoz}, {Bazzon}, {Beuzit}, {Blanchard}, {Brems},
  {Buey}, {De Caprio}, {Carbillet}, {Carle}, {Cascone}, {Cheetham}, {Claudi},
  {Costille}, {Delboulb{\'e}}, {Dohlen}, {Fantinel}, {Feautrier}, {Fusco},
  {Giro}, {Gluck}, {Gry}, {Hubin}, {Hugot}, {Jaquet}, {Le Mignant}, {Llored},
  {Madec}, {Magnard}, {Martinez}, {Maurel}, {Meyer}, {M{\"o}ller-Nilsson},
  {Moulin}, {Mugnier}, {Orign{\'e}}, {Pavlov}, {Perret}, {Petit}, {Pragt},
  {Puget}, {Rabou}, {Ramos}, {Rigal}, {Rochat}, {Roelfsema}, {Rousset}, {Roux},
  {Salasnich}, {Sauvage}, {Sevin}, {Soenke}, {Stadler}, {Suarez}, {Turatto}, \&
  {Weber}}]{Keppler2018}
{Keppler}, M., {Benisty}, M., {M{\"u}ller}, A., {et~al.} 2018, \aap, 617, A44

\bibitem[{{Kuhn} {et~al.}(2001){Kuhn}, {Potter}, \& {Parise}}]{Kuhn2001}
{Kuhn}, J.~R., {Potter}, D., \& {Parise}, B. 2001, \apjl, 553, L189,
  \dodoi{10.1086/320686}

\bibitem[{{Lafreni{\`e}re} {et~al.}(2007){Lafreni{\`e}re}, {Marois}, {Doyon},
  {Nadeau}, \& {Artigau}}]{Lafreniere2007}
{Lafreni{\`e}re}, D., {Marois}, C., {Doyon}, R., {Nadeau}, D., \& {Artigau},
  {\'E}. 2007, \apj, 660, 770, \dodoi{10.1086/513180}

\bibitem[{{Lawson} {et~al.}(2020){Lawson}, {Currie}, {Wisniewski}, {Tamura},
  {Schneider}, {Augereau}, {Brandt}, {Guyon}, {Kasdin}, {Groff}, {Lozi},
  {Chilcote}, {Hodapp}, {Jovanovic}, {Martinache}, {Skaf}, {Akiyama},
  {Henning}, {Knapp}, {Kwon}, {Mayama}, {McElwain}, {Sitko}, {Asensio-Torres},
  {Uyama}, \& {Wagner}}]{Lawson2020}
{Lawson}, K., {Currie}, T., {Wisniewski}, J.~P., {et~al.} 2020, \aj, 160, 163,
  \dodoi{10.3847/1538-3881/ababa6}

\bibitem[{{Lawson} {et~al.}(2021{\natexlab{a}}){Lawson}, {Currie},
  {Wisniewski}, {Tamura}, {Augereau}, {Brandt}, {Guyon}, {Kasdin}, {Groff},
  {Lozi}, {Deo}, {Vievard}, {Chilcote}, {Jovanovic}, {Martinache}, {Skaf},
  {Henning}, {Knapp}, {Kwon}, {McElwain}, {Pyo}, {Sitko}, {Uyama}, \&
  {Wagner}}]{Lawson2021b}
---. 2021{\natexlab{a}}, \aj, 162, 293, \dodoi{10.3847/1538-3881/ac2823}

\bibitem[{{Lawson} {et~al.}(2021{\natexlab{b}}){Lawson}, {Currie},
  {Wisniewski}, {Hashimoto}, {Guyon}, {Kasdin}, {Groff}, {Lozi}, {Brandt},
  {Chilcote}, {Deo}, {Uyama}, \& {Vievard}}]{Lawson2021spie}
{Lawson}, K., {Currie}, T., {Wisniewski}, J.~P., {et~al.} 2021{\natexlab{b}},
  in Society of Photo-Optical Instrumentation Engineers (SPIE) Conference
  Series, Vol. 11823, Society of Photo-Optical Instrumentation Engineers (SPIE)
  Conference Series, 118230D, \dodoi{10.1117/12.2594819}

\bibitem[{{Lomax} {et~al.}(2016){Lomax}, {Wisniewski}, {Grady}, {McElwain},
  {Hashimoto}, {Kudo}, {Kusakabe}, {Okamoto}, {Fukagawa}, {Abe}, {Brandner},
  {Brandt}, {Carson}, {Currie}, {Egner}, {Feldt}, {Goto}, {Guyon}, {Hayano},
  {Hayashi}, {Hayashi}, {Henning}, {Hodapp}, {Inoue}, {Ishii}, {Iye}, {Janson},
  {Kandori}, {Knapp}, {Kuzuhara}, {Kwon}, {Matsuo}, {Mayama}, {Miyama},
  {Momose}, {Morino}, {Moro-Martin}, {Nishimura}, {Pyo}, {Schneider},
  {Serabyn}, {Sitko}, {Suenaga}, {Suto}, {Suzuki}, {Takahashi}, {Takami},
  {Takato}, {Terada}, {Thalmann}, {Tomono}, {Turner}, {Watanabe}, {Yamada},
  {Takami}, {Usuda}, \& {Tamura}}]{Lomax2016}
{Lomax}, J.~R., {Wisniewski}, J.~P., {Grady}, C.~A., {et~al.} 2016, \apj, 828,
  2, \dodoi{10.3847/0004-637X/828/1/2}

\bibitem[{{Lovell} {et~al.}(2021){Lovell}, {Marino}, {Wyatt}, {Kennedy},
  {MacGregor}, {Stapelfeldt}, {Dent}, {Krist}, {Matr{\`a}}, {Kral},
  {Pani{\'c}}, {Pearce}, \& {Wilner}}]{Lovell2021}
{Lovell}, J.~B., {Marino}, S., {Wyatt}, M.~C., {et~al.} 2021, \mnras, 506,
  1978, \dodoi{10.1093/mnras/stab1678}

\bibitem[{{Macintosh} {et~al.}(2015){Macintosh}, {Graham}, {Barman}, {De Rosa},
  {Konopacky}, {Marley}, {Marois}, {Nielsen}, {Pueyo}, {Rajan}, {Rameau},
  {Saumon}, {Wang}, {Patience}, {Ammons}, {Arriaga}, {Artigau}, {Beckwith},
  {Brewster}, {Bruzzone}, {Bulger}, {Burningham}, {Burrows}, {Chen}, {Chiang},
  {Chilcote}, {Dawson}, {Dong}, {Doyon}, {Draper}, {Duch{\^e}ne}, {Esposito},
  {Fabrycky}, {Fitzgerald}, {Follette}, {Fortney}, {Gerard}, {Goodsell},
  {Greenbaum}, {Hibon}, {Hinkley}, {Cotten}, {Hung}, {Ingraham},
  {Johnson-Groh}, {Kalas}, {Lafreniere}, {Larkin}, {Lee}, {Line}, {Long},
  {Maire}, {Marchis}, {Matthews}, {Max}, {Metchev}, {Millar-Blanchaer},
  {Mittal}, {Morley}, {Morzinski}, {Murray-Clay}, {Oppenheimer}, {Palmer},
  {Patel}, {Perrin}, {Poyneer}, {Rafikov}, {Rantakyr{\"o}}, {Rice}, {Rojo},
  {Rudy}, {Ruffio}, {Ruiz}, {Sadakuni}, {Saddlemyer}, {Salama}, {Savransky},
  {Schneider}, {Sivaramakrishnan}, {Song}, {Soummer}, {Thomas}, {Vasisht},
  {Wallace}, {Ward-Duong}, {Wiktorowicz}, {Wolff}, \&
  {Zuckerman}}]{Macintosh2015}
{Macintosh}, B., {Graham}, J.~R., {Barman}, T., {et~al.} 2015, Science, 350,
  64, \dodoi{10.1126/science.aac5891}

\bibitem[{{Males} {et~al.}(2018){Males}, {Close}, {Miller}, {Schatz},
  {Doelman}, {Lumbres}, {Snik}, {Rodack}, {Knight}, {Van Gorkom}, {Long},
  {Hedglen}, {Kautz}, {Jovanovic}, {Morzinski}, {Guyon}, {Douglas}, {Follette},
  {Lozi}, {Bohlman}, {Durney}, {Gasho}, {Hinz}, {Ireland}, {Jean}, {Keller},
  {Kenworthy}, {Mazin}, {Noenickx}, {Alfred}, {Perez}, {Sanchez}, {Sauve},
  {Weinberger}, \& {Conrad}}]{Males2018}
{Males}, J.~R., {Close}, L.~M., {Miller}, K., {et~al.} 2018, in Society of
  Photo-Optical Instrumentation Engineers (SPIE) Conference Series, Vol. 10703,
  Adaptive Optics Systems VI, ed. L.~M. {Close}, L.~{Schreiber}, \&
  D.~{Schmidt}, 1070309, \dodoi{10.1117/12.2312992}

\bibitem[{{Marois} {et~al.}(2014){Marois}, {Correia}, {Galicher}, {Ingraham},
  {Macintosh}, {Currie}, \& {De Rosa}}]{Marois2014}
{Marois}, C., {Correia}, C., {Galicher}, R., {et~al.} 2014, in Society of
  Photo-Optical Instrumentation Engineers (SPIE) Conference Series, Vol. 9148,
  Adaptive Optics Systems IV, ed. E.~{Marchetti}, L.~M. {Close}, \& J.-P.
  {Vran}, 91480U, \dodoi{10.1117/12.2055245}

\bibitem[{{Marois} {et~al.}(2010){Marois}, {Macintosh}, \&
  {V{\'e}ran}}]{Marois2010a}
{Marois}, C., {Macintosh}, B., \& {V{\'e}ran}, J.-P. 2010, in Society of
  Photo-Optical Instrumentation Engineers (SPIE) Conference Series, Vol. 7736,
  Adaptive Optics Systems II, ed. B.~L. {Ellerbroek}, M.~{Hart}, N.~{Hubin}, \&
  P.~L. {Wizinowich}, 77361J, \dodoi{10.1117/12.857225}

\bibitem[{{Mawet} {et~al.}(2014){Mawet}, {Milli}, {Wahhaj}, {Pelat}, {Absil},
  {Delacroix}, {Boccaletti}, {Kasper}, {Kenworthy}, {Marois}, {Mennesson}, \&
  {Pueyo}}]{Mawet2014}
{Mawet}, D., {Milli}, J., {Wahhaj}, Z., {et~al.} 2014, \apj, 792, 97,
  \dodoi{10.1088/0004-637X/792/2/97}

\bibitem[{{Mazoyer} {et~al.}(2020){Mazoyer}, {Arriaga}, {Hom},
  {Millar-Blanchaer}, {Chen}, {Wang}, {Duch{\^e}ne}, {Patience}, \&
  {Pueyo}}]{Mazoyer2020}
{Mazoyer}, J., {Arriaga}, P., {Hom}, J., {et~al.} 2020, in Society of
  Photo-Optical Instrumentation Engineers (SPIE) Conference Series, Vol. 11447,
  Society of Photo-Optical Instrumentation Engineers (SPIE) Conference Series,
  1144759, \dodoi{10.1117/12.2560091}

\bibitem[{Newville {et~al.}(2014)Newville, Stensitzki, Allen, \&
  Ingargiola}]{Newville2014}
Newville, M., Stensitzki, T., Allen, D.~B., \& Ingargiola, A. 2014, LMFIT:
  Non-Linear Least-Square Minimization and Curve-Fitting for Python,  Zenodo,
  \dodoi{10.5281/ZENODO.11813}

\bibitem[{Okuta {et~al.}(2017)Okuta, Unno, Nishino, Hido, \&
  Crissman}]{cupy2017}
Okuta, R., Unno, Y., Nishino, D., Hido, S., \& Crissman. 2017, in 31st
  Conference on Neural Information Processing Systems

\bibitem[{{Pairet} {et~al.}(2021){Pairet}, {Cantalloube}, \&
  {Jacques}}]{Pairet2021}
{Pairet}, B., {Cantalloube}, F., \& {Jacques}, L. 2021, \mnras, 503, 3724,
  \dodoi{10.1093/mnras/stab607}

\bibitem[{Perrin {et~al.}(2014)Perrin, Sivaramakrishnan, Lajoie, Elliott,
  Pueyo, Ravindranath, \& Albert}]{Perrin2014}
Perrin, M.~D., Sivaramakrishnan, A., Lajoie, C.-P., {et~al.} 2014, in Space
  Telescopes and Instrumentation 2014: Optical, Infrared, and Millimeter Wave,
  ed. J.~M.~O. Jr., M.~Clampin, G.~G. Fazio, \& H.~A. MacEwen, Vol. 9143,
  International Society for Optics and Photonics (SPIE), 1174 -- 1184,
  \dodoi{10.1117/12.2056689}

\bibitem[{{Pueyo}(2016)}]{Pueyo2016}
{Pueyo}, L. 2016, \apj, 824, 117, \dodoi{10.3847/0004-637X/824/2/117}

\bibitem[{{Ren} {et~al.}(2020){Ren}, {Pueyo}, {Chen}, {Choquet}, {Debes},
  {Duch{\^e}ne}, {M{\'e}nard}, \& {Perrin}}]{Ren2020}
{Ren}, B., {Pueyo}, L., {Chen}, C., {et~al.} 2020, \apj, 892, 74,
  \dodoi{10.3847/1538-4357/ab7024}

\bibitem[{{Ren} {et~al.}(2018){Ren}, {Pueyo}, {Zhu}, {Debes}, \&
  {Duch{\^e}ne}}]{Ren2018}
{Ren}, B., {Pueyo}, L., {Zhu}, G.~B., {Debes}, J., \& {Duch{\^e}ne}, G. 2018,
  \apj, 852, 104, \dodoi{10.3847/1538-4357/aaa1f2}

\bibitem[{{Rich} {et~al.}(2022){Rich}, {Monnier}, {Aarnio}, {Laws},
  {Setterholm}, {Wilner}, {Calvet}, {Harries}, {Miller}, {Davies}, {Adams},
  {Andrews}, {Bae}, {Espaillat}, {Greenbaum}, {Hinkley}, {Kraus}, {Hartmann},
  {Isella}, {McClure}, {Oppenheimer}, {P{\'e}rez}, \& {Zhu}}]{Rich2022}
{Rich}, E.~A., {Monnier}, J.~D., {Aarnio}, A., {et~al.} 2022, arXiv e-prints,
  arXiv:2206.05815.
\newblock \doarXiv{2206.05815}

\bibitem[{{Soummer} {et~al.}(2012){Soummer}, {Pueyo}, \&
  {Larkin}}]{Soummer2012}
{Soummer}, R., {Pueyo}, L., \& {Larkin}, J. 2012, \apjl, 755, L28,
  \dodoi{10.1088/2041-8205/755/2/L28}

\bibitem[{{Stolker} {et~al.}(2016){Stolker}, {Dominik}, {Min}, {Garufi},
  {Mulders}, \& {Avenhaus}}]{Stolker2016}
{Stolker}, T., {Dominik}, C., {Min}, M., {et~al.} 2016, \aap, 596, A70,
  \dodoi{10.1051/0004-6361/201629098}

\bibitem[{{Takami} {et~al.}(2013){Takami}, {Karr}, {Hashimoto}, {Kim},
  {Wisniewski}, {Henning}, {Grady}, {Kandori}, {Hodapp}, {Kudo}, {Kusakabe},
  {Chou}, {Itoh}, {Momose}, {Mayama}, {Currie}, {Follette}, {Kwon}, {Abe},
  {Brandner}, {Brandt}, {Carson}, {Egner}, {Feldt}, {Guyon}, {Hayano},
  {Hayashi}, {Hayashi}, {Ishii}, {Iye}, {Janson}, {Knapp}, {Kuzuhara},
  {McElwain}, {Matsuo}, {Miyama}, {Morino}, {Moro-Martin}, {Nishimura}, {Pyo},
  {Serabyn}, {Suto}, {Suzuki}, {Takato}, {Terada}, {Thalmann}, {Tomono},
  {Turner}, {Watanabe}, {Yamada}, {Takami}, {Usuda}, \& {Tamura}}]{Takami2013}
{Takami}, M., {Karr}, J.~L., {Hashimoto}, J., {et~al.} 2013, \apj, 772, 145,
  \dodoi{10.1088/0004-637X/772/2/145}

\bibitem[{{Tazaki} \& {Dominik}(2022)}]{Tazaki2022}
{Tazaki}, R., \& {Dominik}, C. 2022, \aap, 663, A57,
  \dodoi{10.1051/0004-6361/202243485}

\bibitem[{{Thalmann} {et~al.}(2013){Thalmann}, {Janson}, {Buenzli}, {Brandt},
  {Wisniewski}, {Dominik}, {Carson}, {McElwain}, {Currie}, {Knapp},
  {Moro-Mart{\'\i}n}, {Usuda}, {Abe}, {Brandner}, {Egner}, {Feldt}, {Golota},
  {Goto}, {Guyon}, {Hashimoto}, {Hayano}, {Hayashi}, {Hayashi}, {Henning},
  {Hodapp}, {Ishii}, {Iye}, {Kandori}, {Kudo}, {Kusakabe}, {Kuzuhara}, {Kwon},
  {Matsuo}, {Mayama}, {Miyama}, {Morino}, {Nishimura}, {Pyo}, {Serabyn},
  {Suto}, {Suzuki}, {Takami}, {Takato}, {Terada}, {Tomono}, {Turner},
  {Watanabe}, {Yamada}, {Takami}, \& {Tamura}}]{Thalmann2013}
{Thalmann}, C., {Janson}, M., {Buenzli}, E., {et~al.} 2013, \apjl, 763, L29,
  \dodoi{10.1088/2041-8205/763/2/L29}

\bibitem[{{van der Marel} {et~al.}(2021){van der Marel}, {Birnstiel}, {Garufi},
  {Ragusa}, {Christiaens}, {Price}, {Sallum}, {Muley}, {Francis}, \&
  {Dong}}]{vanderMarel2021}
{van der Marel}, N., {Birnstiel}, T., {Garufi}, A., {et~al.} 2021, \aj, 161,
  33, \dodoi{10.3847/1538-3881/abc3ba}

\bibitem[{Virtanen {et~al.}(2020)Virtanen, Gommers, Oliphant, Haberland, Reddy,
  Cournapeau, Burovski, Peterson, Weckesser, Bright, {van der Walt}, Brett,
  Wilson, Millman, Mayorov, Nelson, Jones, Kern, Larson, Carey, Polat, Feng,
  Moore, {VanderPlas}, Laxalde, Perktold, Cimrman, Henriksen, Quintero, Harris,
  Archibald, Ribeiro, Pedregosa, {van Mulbregt}, \& {SciPy 1.0
  Contributors}}]{scipy2020}
Virtanen, P., Gommers, R., Oliphant, T.~E., {et~al.} 2020, Nature Methods, 17,
  261, \dodoi{10.1038/s41592-019-0686-2}

\bibitem[{{Whitney} {et~al.}(2013){Whitney}, {Robitaille}, {Bjorkman}, {Dong},
  {Wolff}, {Wood}, \& {Honor}}]{Whitney2013}
{Whitney}, B.~A., {Robitaille}, T.~P., {Bjorkman}, J.~E., {et~al.} 2013, \apjs,
  207, 30, \dodoi{10.1088/0067-0049/207/2/30}

\end{thebibliography}
\bibliographystyle{aasjournal}
\end{document}